\documentclass[onecolumn]{aastex631}
\usepackage{amsmath}
\usepackage{bm}

\shorttitle{FRB lensing}
\shortauthors{Connor \& Ravi}
\graphicspath{{./}{figures/}}

\begin{document}

\title{Stellar prospects for FRB gravitational lensing}

\author{Liam Connor}
\affiliation{Cahill Center for Astronomy and Astrophysics, MC 249-17 California Institute of Technology, Pasadena, CA 91125, USA}
\correspondingauthor{Liam Connor}
\email{liam.dean.connor@gmail.com}

\author{Vikram Ravi}
\affiliation{Cahill Center for Astronomy and Astrophysics, MC 249-17 California Institute of Technology, Pasadena, CA 91125, USA}

\begin{abstract}
Gravitational lensing of fast radio bursts (FRBs) offers 
an exciting avenue for several cosmological applications. 
However, it is not yet clear how many such events 
future surveys will detect nor how to optimally find 
them.
We use the known properties of FRBs to 
forecast detection rates of gravitational lensing on delay timescales from microseconds to years, corresponding to lens masses spanning fifteen orders of magnitude. 
We highlight the role of the FRB redshift distribution 
on our ability to observe gravitational lensing. We consider 
cosmological lensing of FRBs by stars  
in foreground galaxies and show that strong stellar lensing 
will dominate on microsecond timescales. 
Upcoming surveys such as DSA-2000 and CHORD will constrain the fraction of dark matter 
in compact objects (e.g. primordial black holes) and may 
detect millilensing events from intermediate mass black holes (IMBHs) or small dark matter halos. Coherent all-sky monitors 
will be able to detect longer-duration  
lensing events from massive galaxies, in addition to short time-scale lensing. Finally, we propose a new application of FRB gravitational lensing that will measure directly the circumgalactic medium of intervening galaxies. 

\end{abstract}


\section{Introduction}
\label{sec:intro}
Gravitational lensing is the deflection 
of light rays by intervening matter inhomogeneities 
between a source and an observer. If the deflection angle 
is sufficiently large, one can observe multiple images 
of the source. Each image will traverse a different path, 
leading to arrival time delays between the lensed copies. 
Therefore, variable 
sources and astrophysical transients allow one to detect 
gravitational lensing in the time domain. The differential 
arrival time of lensed images enable 
valuable cosmological applications, for example measuring 
the Hubble constant $H_0$ with a technique 
known as time-delay cosmography \citep{refsdal, cosmography-review-Treu, holicow}. 

Gravitational lensing in the time domain has primarily been observed 
in distant quasars \citep{Vanderriest-1989, cosmography-review-Treu}, the brightness of which  
fluctuates on human timescales due to their compact emitting regions. 
There are also several gravitationally lensed 
events or candidates from explosive transients \citep{oguri-2019} such as supernovae \citep{Kelly-2015, Rodney-2021}, 
gamma-ray bursts (GRBs) \citep{Paynter-2022}, and gravitational waves 
events from coalescing binary black holes.

\begin{figure}[htb]
    \centering
    \includegraphics[width=0.8\columnwidth]{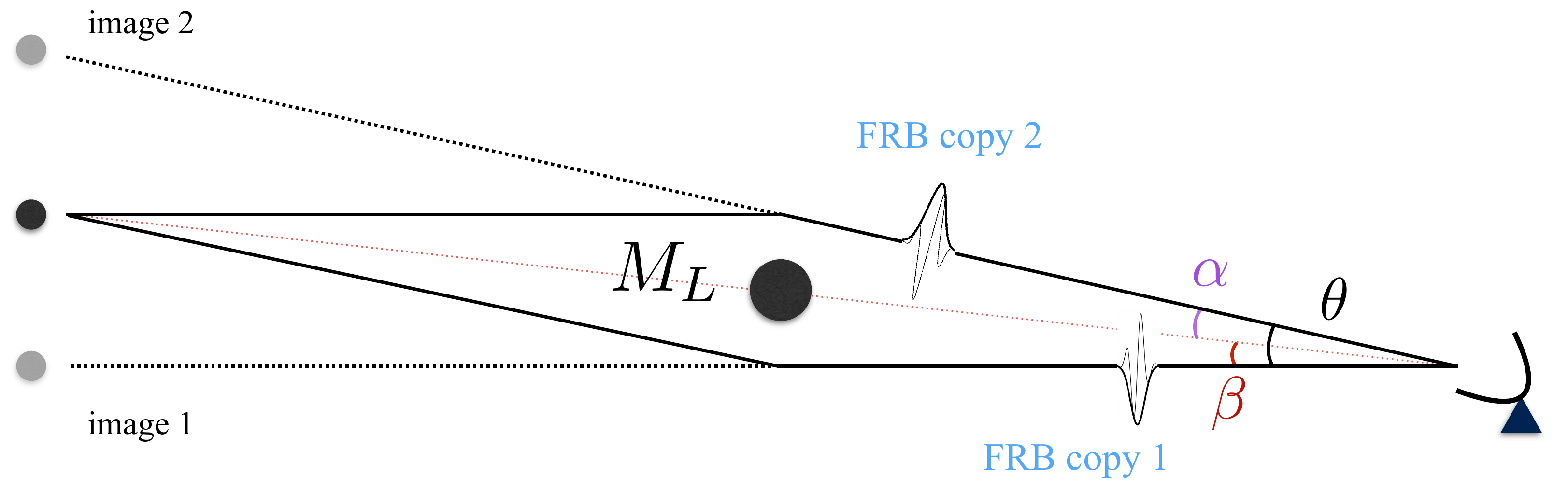}
    \caption{Diagram of fast radio burst gravitational lensing.}
    \label{fig:lensing_delay}
\end{figure}

Fast radio bursts (FRBs) are bright, brief radio transients 
whose exact origins remain unknown \citep{petroff-2019, cordes-review}. 
FRBs offer a uniquely precise 
probe of gravitational lensing in the time domain for two reasons: 
They are ubiquitous, with volumetric rates that may exceed 
those of core-collapse supernovae. And they are very short in duration, 
allowing for extraordinary measurements of the lensing time-delay. Typical bursts are $\sim$\,millisecond duration, but radio telescopes can preserve 
electric field information about the FRB on timescales of 
nanoseconds, well below the already-narrow pulse widths. This is in contrast to other time-delay lensing 
events, which are fundamentally limited in the delays---and 
therefore lens mass-scales---to which they are sensitive. In gravitational lensing, the image that arrives first 
will be brighter than subsequent copies. 
A schematic diagram of FRB lensing is shown in 
Figure~\ref{fig:lensing_delay}. 

\begin{figure}
    \centering
    \includegraphics[width=0.85\columnwidth]{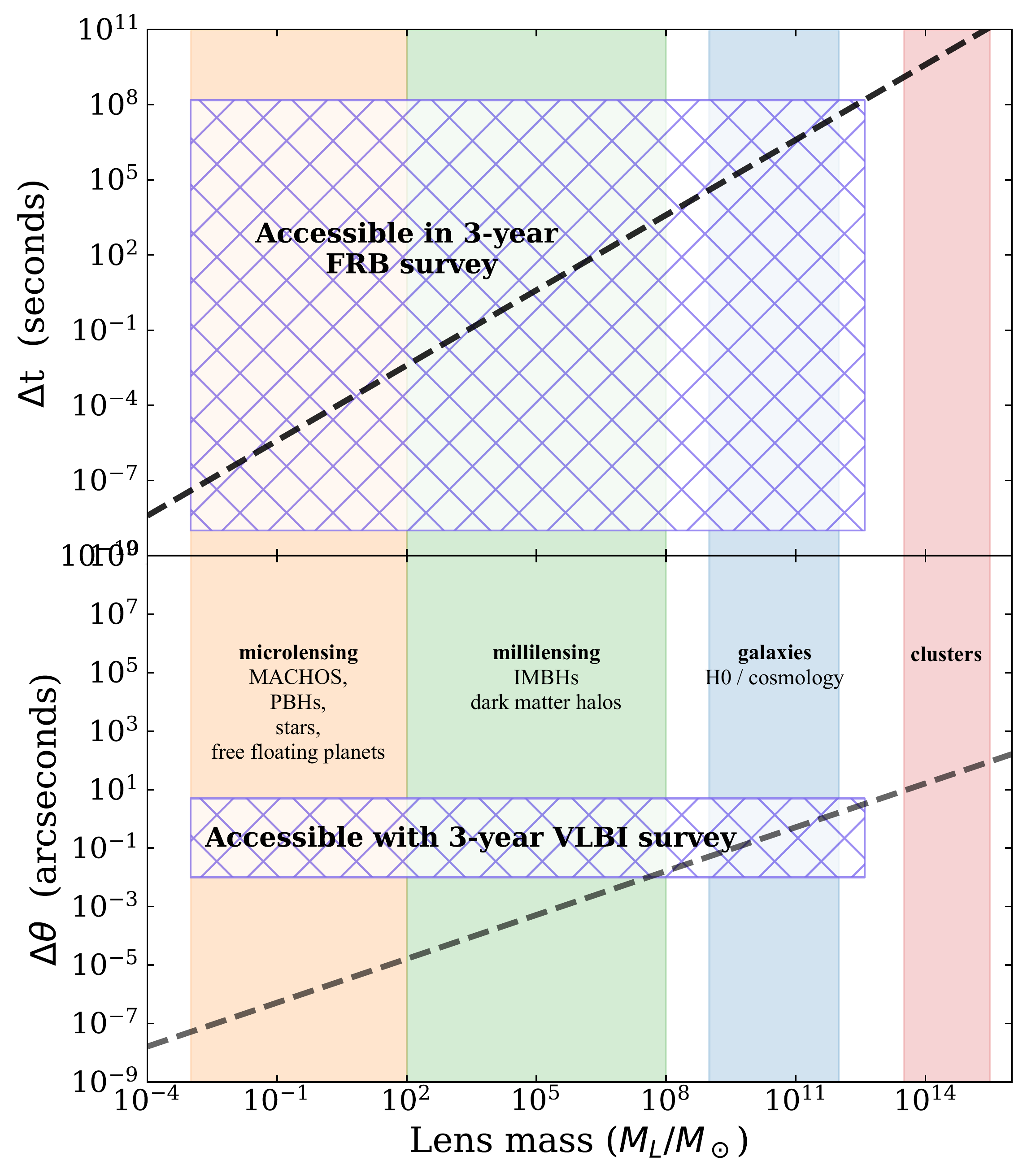}
    \caption{The phase space of FRB gravitational 
    lensing for a source at $z=1$ and lens at $z=0.5$.
    The dashed black curves in the top and bottom 
    panels show the fiducial time delay and 
    image separation, respectively, as a function of lens mass. The solid shaded regions give examples of 
    the types of lenses at each mass scale.}
    \label{fig:phase_space}
\end{figure}

In the past several years, a number of groups have explored 
the application of FRB gravitational lensing to cosmology and fundamental 
physics \citep{Zheng-2014, li2018, Wucknitz2021, chen-frb-lens-I}. \citet{munos2016} outlined how searching for 
FRBs lensed on timescales of
milliseconds would constrain the 
fraction of dark matter in massive compact halo objects 
(MACHOS), particularly in the mass range 20--100\,M$_\odot$. 
Relatedly, FRBs have been proposed as a probe of the 
mass distribution function of primordial black holes (PBHs)
between 10 and 10$^3$\,M$_\odot$ \citep{zhou-PBH}.
These methods are incoherent in that phase information about the 
electric field (i.e. the burst's recorded complex voltage data) 
was not considered. 
Such methods have an identification problem because 
many FRBs repeat, so one must distinguish a genuine lensing 
event from a distinct burst from the same source. However, if
voltage data---rather than just total intensity---of the 
burst are preserved, lensing delays can theoretically be detected 
down to the instrument's inverse radio bandwidth ($B$\,$\sim$\,1\,GHz or $\Delta t$\,$\sim$\,10$^{-9}$\,s) \citep{Eichler2017, pen18}. Barring 
instrumental and propagation effects, lensed copies of the same 
burst ought to have identical waveforms. The same will not be true 
for intrinsically different bursts, whether from a repeating 
source or from a different FRB source along a similar sight line. The unprecedented access to time delays from nanoseconds to years means FRBs could probe lens mass scales from Jupiter-like objects up to massive galaxies, spanning many orders of magnitude in 
lens masses. Indeed, techniques for coherent time-delay gravitational 
lensing of compact sources such as pulsars and FRBs 
have been proposed for lensing by free-floating planets in the Milky Way \citep{jow-2020}. For longer 
time delays, propagation effects due to plasma, such as 
scattering, have deleterious effects on the coherent lensing signal. 
In those cases, being able to spatially resolve the multiple images using very-long baseline interferoemtry (VLBI) will enable new science for lenses that are galaxy-scale and above. The large range 
of time delays, image separations, and lens masses 
accessible to FRBs is shown in Figure~\ref{fig:phase_space}.

Recently, these ideas have been developed more extensively 
and put into practice by the CHIME/FRB collaboration 
\citep{kader-2022, leung-2022}. The authors 
have built coherent methods 
for searching voltage data of CHIME/FRB sources for lensing events, which they refer to as ``gravitational lens interferometry''. They are able to search for lensing delays between 2.5\,ns and 100\,ms, corresponding to $10^{-4}$--$10^{4}$\,M$_\odot$ lenses. They have applied these techniques to 
CHIME/FRB data to 
constrain the fraction of dark matter in PBHs using 
172 bursts for which voltage data was preserved, 
accounting for decoherence from scattering \citep{leung-2022}. 
This amounted to 114 distinct sightlines, meaning a positive detection would have required a very high cosmic abundance of PBHs. 

Despite the clear value of having a collection of gravitationally 
lensed FRBs from cosmological distances, it remains an 
open question how future surveys should optimally search for them or how 
many they will detect. The purpose of this work 
is to compare different telescope designs and to produce a realistic 
forecast for how many lensed FRBs future and current experiments might find. This allows us to discuss several science cases at a wide range of lensing timescales (and thus lens masses). We first offer a basic formalism for gravitational 
lensing in the time domain and calculate lensing optical 
depths at different mass scales. We then forecast total FRB 
detection rates on DSA-2000 \citep{dsa-2000-whitepaper}, CHORD \citep{Vanderlinde-2019}, and a coherent all-sky monitor (for example, BURSTT \citep{burstt-2022}\footnote{https://www.asiaa.sinica.edu.tw/project/burstt.php}), finding they will detect tens of thousands of new FRBs. We combine these rates and their modelled redshift distributions with lensing optical depths to estimate the number 
of gravitationally lensed FRBs that will be detected on a range of timescales. 
We describe the science that can be done with lens masses of $10^{-1}$--$10^{14}$\,$M_\odot$.
Finally we describe the application of FRB lensing to the circumgalactic medium (CGM), 
providing a clean measurement of halo gas properties along different lines of sight.

\section{Gravitational lensing background}

If a source is at a true sky position, $\bm{\beta}$, 
its image will appear at $\bm{\theta}$, deflected 
by the angle $\bm{\alpha}$ in the presence of a gravitational 
lens. The mapping between the true and apparent source 
position is given by the lens equation, 

\begin{equation}
    \boldmath{\beta} =  \mathbf{\theta} -  \mathbf{\alpha}( \mathbf{\theta} ).
    \label{eq:lens}
\end{equation}

\noindent The deflection angle, $\bm{\alpha}$, 
is given by an integral of the projected 
surface mass density, $\Sigma(\bm{\theta})$ over 
angular position,

\begin{equation}
    \bm{\alpha}( \bm{\theta}) = \frac{1}{\pi} \int 
    \mathrm{d}\bm{\theta'} \frac{|\bm{\theta}-\bm{\theta'}|}{|\bm{\theta}-\bm{\theta'}|^2}
    \frac{\Sigma(\bm{\theta'})}{\Sigma_{cr}}.
\end{equation}

\noindent Here $\Sigma_{cr}$ is the critical surface density and 
is given by the geometry of the lensing system, 

\begin{equation}
    \Sigma_{cr} = \frac{c^2}{4\pi\,G} \frac{D_s}{D_l\,D_{ls}},
\end{equation}

\noindent where $D_s$, $D_l$, and $D_{ls}$ are the distances to the source, 
the lens, and the distance between the lens and the source, 
respectively. $\Sigma_{cr}$, and therefore the solution to 
the lens equation, also depend on the constituents of the Universe 
because $D_s$, $D_l$, and $D_{ls}$ are all angular diameter distances. 

The Lens Equation is non-linear in 
$\bm{\theta}$, so there can be more than 
one $\bm{\theta}$ that satisfies Eq~\ref{eq:lens}.
Hence, gravitational lenses produce multiple images 
or multiple lensed copies in time. A special case 
is when the source is directly behind the lens where 
$\bm{\beta}=0$. In that scenario, the deflection angle 
is equal to the image position. That angle is known as the Einstein radius, $\theta_E = \alpha$.

The next useful quantity for our purposes is the lensing 
optical depth, which is the probability that a source 
at redshift $z_s$ is lensed. It will be an integral 
of the cross sections ($\sigma\approx\pi\,\theta^2_E$) of all lenses between the observer and the source. 

\begin{equation}
\tau(z_s) = \int_0^{z_s}\mathrm{d}z_l\int n(\sigma, z_l)
\,\sigma
\frac{d^2V}{d\Omega dz_l}
\end{equation}

Here, $n(\sigma, z_l)$ is the number density of sources 
at redshift with cross-sectional area $\sigma$ and $\frac{d^2V}{d\Omega dz_l}$ 
is the comoving volume element per redshift per 
steradian. 

We are concerned with the rate of 
FRB lensing events, for which we must incorporate the 
source redshift distribution. Assuming an FRB detection 
rate per redshift (in the absence of lensing) of $\mathcal{R}_{det}(z_s)$, the 
rate of observed lensing events, $\mathcal{R}_L$, will be the following:

\begin{equation}
\mathcal{R}_L = \int^\infty_0 dz_s \, \mathcal{R}_{det}(z_s) \int_0^{z_s}\mathrm{d}z_l\int\,B(\gamma)\, n(\sigma, z_l)
\,\sigma
\frac{d^2V}{d\Omega dz_l}.
\label{eq:lens-rate-eq}
\end{equation}

The parameter $B$ accounts for a phenomenon known as 
magnification bias \citep{turner-1980}. Since gravitational lensing can 
magnify the intensity of lensed copies, faint sources 
that are otherwise below a survey's detection threshold 
can be made observable. This increases the number of 
lensed objects and must be accounted for when computing 
optical depth. Assuming a power-law luminosity function, 
$N(L)\propto L^{-\gamma}$, and a singular isothermal sphere (SIS) lens, 

\begin{equation}
    B(\gamma) = \frac{2^\gamma}{3-\gamma}.
\label{eq:gamma}
\end{equation}

\noindent The equation shows magnification bias 
is stronger for steeper luminosity functions, 
i.e. large $\gamma$. This is because steep 
power-laws indicate an abundance of 
low-luminosity sources that can only be 
observed in the presence of lensing magnification for a flux-limited survey. 

Evidently, the main ingredients that impact 
the detection of FRB lensing will be
the abundance of lenses, the mass distribution of those lenses, and the redshift 
distribution of FRBs that a given survey observes. 
Another important factor for transient lensing will be sky coverage and observing strategy, 
which is captured by the ``time-delay selection function''.

\subsubsection{Time-delay selection function}
In the image-domain, only one image is needed 
to search for evidence of gravitational lensing.
But for time-domain events such as FRBs, GRBs, SNe, or time-variable AGN, one must be 
pointing at the same patch of sky when the 
lensed signal arrives. This poses a challenge 
for transient gravitational lensing. We call the probability that the lensed copy will be in the telescope's field-of-view (FoV) $P_{\rm FoV}(\Delta t)$. This determines the likelihood 
that a lensed image will be recorded by a time-domain survey.
When calculating the true detection rate of lensed FRBs with Eq~\ref{eq:lens-rate-eq}, we must multiply the integrand by $P_{\rm FoV}(\Delta t)$.

For a transit instrument with east west beamwidth, $\theta_{\rm EW}$, lensed copies will be recorded if the delay is less than a beam crossing time, $t_{\rm trans}$. If the delay 
is greater than a transit time but less than one day, it cannot be detected. For delays longer than one day, the probability of detecting the lensed copy is the fractional sky coverage, or $\frac{\theta_{\rm EW
}}{2\pi}$. Examples of such FRB surveys are CHIME/FRB \citep{chime-frb-overview}, DSA-110\footnote{https://www.deepsynoptic.org/team-110}, and CHORD \citep{Vanderlinde-2019}. The probability that the lensed copy will be in the beam is therefore,

\begin{equation}
P_{\rm FoV}(\Delta t) = \left\{\begin{matrix}
1 &  \Delta t \leq t_{\rm trans} \\ 
0 &   t_{\rm trans} < \Delta t < 1\,\rm day \\ 
\frac{\theta_{\rm EW
}}{2\pi} & \Delta t \geq 1\,\rm day 
\end{matrix}\right.
\end{equation}

\noindent For CHORD, $P_{\rm FoV}(\Delta t\geq10^5\,\mathrm{s})\approx9\times10^{-3}$ at the center of the band, assuming the instrument is parked at the same declination. For a steerable all-sky survey such as DSA-2000, 
$P_{\rm FoV}(\Delta t)=1$ for $\Delta t$ less than a 
pointing time. For longer time-delays, $P_{\rm FoV}(\Delta t)$
will be roughly the fraction of time spent on each patch of sky, 
or the primary beam size divided by $3\,\pi$ ($P_{\rm FoV}(\Delta t\geq10^3\,\mathrm{s})\approx3\times10^{-4}$ on DSA-2000). 
In practice, CHIME/FRB, CHORD, DSA-110, and DSA-2000 
will have a difficult time detecting gravitational lensing events caused by massive 
galaxies unless special survey strategy is undertaken. Instead, they will be able to search for 
lensing by compact objects and halos with $M_L\lesssim10^8\,M_\odot$.

An ultra-widefield FRB survey, which observes a similar region of sky at all times, will be able to search for 
lensing delays up to the survey duration. We refer to such surveys as 
coherent all-sky monitors, or CASMs, of which the proposed experiment 
BURSTT is an example \citep{burstt-2022}. 
However, such a design 
will inevitably be less sensitive than DSA-2000 
or CHORD and will therefore probe a more nearby population of FRBs. 
In Section~\ref{section:results} we compare this trade-off. Another way to 
access longer lensing time-delays is to point one's telescope 
at the north or south celestial pole such that the same 
field is observed at all times or observe a circumpolar region 
with a tracking telescope.

\subsection{Lensing by a point mass}
We start with the simplest possible mass 
distribution, that of a point mass. We calculate both the Einstein 
radius and typical time-delay for a 
cosmological source. This will be relevant for stellar lenses, 
primordial black holes / MACHOS, and
intermediate-mass black holes. 
For a point mass, 
the Einstein radius scales as the square root of mass
as,

\begin{equation}
        \theta_{E} = \sqrt{\frac{4\,G\,M}{c^2}\,\frac{D_{ls}}{D_l\,D_s}}
\end{equation}

\begin{equation}
    \approx 1.6''\times10^{-6}\left ( \frac{M}{M_\odot}\right)^{1/2} \left ( \frac{D_l\,D_s/D_{ls}}{3\,\mathrm{Gpc}}\right)^{-1/2}
\end{equation}

To compute a time-delay between images, 
we follow \cite{oguri-2019} by defining a 
fiducial time-delay, $\Delta t_{fid}$, which is 
the difference in arrival time between the unlensed line-of-sight and an image deflected by 
$\theta_E$. In practice, the true observable 
between images $i$ and $j$ is the difference between 
their time-delays, i.e. we measure $\Delta t_{ij} = \Delta t_{i} - \Delta t_{j}$. It is still useful to consider 
the typical lensing delay timescale. This is given by,

\begin{equation}
    \Delta t_{fid} \approx 20\,\mu\mathrm{s}\times (1 + z_l) \left ( \frac{M}{M_{\odot}}\right ).
\end{equation}

\noindent The point lens will produce images separated 
by the following,

\begin{equation}
    \theta_\pm = \frac{1}{2}\left (\beta\pm\sqrt{\beta^2 + 4\theta^2_E}\right ),
\end{equation}

\noindent where $\beta$ is angular impact parameter. The time delay will be,

\begin{equation}
    \Delta t = \frac{4\,G M_l}{c^3}\,(1+z_l)\,\left (\frac{y}{2}\sqrt{y^2+4} + \log \left (\frac{\sqrt{y^2+4} + y}{\sqrt{y^2+4} - y} \right ) \right ),
\end{equation}

\noindent where $y\equiv\beta/\theta_E$  \citep{munos2016}. 

In order to calculate the lensing optical depth, one must usually account for the fact that the area is an annulus between $y_{min}$ and 
$y_{max}$ rather than simply a cross section of
$\pi\theta^2_E$. The areal scale of the lens is then,

\begin{equation}
    \sigma(z_l, M_l) = \frac{4\pi\,G\,M_l}{c^2}\,
    \frac{D_l\,D_{ls}}{D_s}\,\left (y^2_{max} - y^2_{min} \right ).
\end{equation}

Here, $y_{max}$ is set by the largest acceptable image pair flux ratio. The minimum value, $y_{min}$,
is set by the minimum detectable 
time delay between the two images. Unlike with GRBs \citep[e.g.]{Paynter-2022} 
or previous incoherent treatments of FRB lensing \citep{munos2016, zhou-PBH}, we assume a coherent 
lensed search that can find delays below the pulse width scale, $\Delta t<< t_{FRB}$ \citep{Eichler2017, Wucknitz2021}. This renders $y_{min}$ negligible 
so we set it to zero going forward. The source number density is,

\begin{equation}
    n(z_l, M_l) = \frac{\rho_c\,f_{l}\,\Omega_c}{M_l} (1+z)^3,
\end{equation}

\noindent where $\rho_c$ is the critical density 
of the Universe, $\Omega_c$ is the cosmological 
density parameter of cold dark matter at $z=0$, and $f_l$ 
is the fraction of dark matter in compact lenses 
(e.g. in primordial black holes or IMBHs). Combining 
these terms, we find that the lensing optical depth 
is independent of the mass of compact halos. It is given by,

\begin{equation}
    \tau(z_s) = \frac{3}{2} f_{l} \Omega_{c} H^2_0
    \int^{z_s}_0 \mathrm{d}z_l\,B\,\frac{(1+z_l)^2}{c\,H(z_l)} \frac{D_l\,D_{ls}}{D_s}
    y^2_{max}.
\end{equation}

As a useful guide, the classic Press-Gunn approximation relates the cosmic abundance of lenses to their mean optical depth as $\tau \approx \Omega_l$
for high redshift sources, where $\Omega_l$ is the cosmological parameter for 
that type of point-mass lens \citep{press73}. For lower redshifts, the scaling is closer 
to $\tau \approx \Omega_l \left ( \frac{z}{2} \right)^2$ \citep{narayan-1996}.
Notably, the relation is independent of lens mass. 

In Figure~\ref{fig:lens_prob} we plot lensing probability curves for two point mass scenarios as a function of 
redshift. The dotted curve assumes 0.1$\%$ of dark matter 
is made from compact halos and the dot-dashed curve 
assumes this number is $10^{-4}$, both of which 
are currently permitted in the relevant lens mass range. 
Each uses a magnification bias of 2.

\subsection{Strong lensing by massive galaxies}
The lensing optical depth for distant objects 
is dominated by the dark matter halos of 
massive foreground galaxies. Lensing halos 
are typically taken to be singular isothermal spheres 
(SIS) or ellipsoids (SIE). These objects are 
parameterized by their Einstein radius given by,

\begin{equation}
    \theta_E = 4\pi\frac{\sigma_v^2}{c^2}\frac{D_{ls}}{D_s},
\end{equation}

\noindent where $\sigma_v$ is the halos' velocity dispersion. 
For reference, with $\sigma_v=200$\,km\,s$^{-1}$ and $D_{ls}/{D_s}$=0.426, the Einstein radius is 
roughly 0.5'', or a typical image separation of 
one arcsecond. The optical depth is then,

\begin{equation}
\tau(z_s) = \int_0^{z_s}\mathrm{d}z_l\int d\sigma_v 
B(\gamma)\,\phi(\sigma_v, z_l)\frac{d^2V}{d\Omega dz_l}
\pi \theta^2_E(\sigma, z_l, z_s) D^2_l,
\end{equation}

\noindent where $\frac{d^2V}{d\Omega dz_l}$ is 
the differential comoving volume at $z_l$ and $\phi(\sigma, z_l)$
is the number density of galaxies with velocity dispersion 
$\sigma$, also known as velocity dispersion function (VDF).

To model the lensing optical depth from 
massive galaxies we use a model based on 
a recent empirical galaxy VDF \citep{Yue-analytical}. Another quick 
approximation for lensing probability is given by \citep{oguri-2019},

\begin{equation}
    \tau(z_s) = \frac{5\times10^{-4}\,z_s^3}{(1+0.041z^{1.1}_s)^{2.7}}
\end{equation}

\begin{figure}[htb]
    \centering
    \includegraphics[width=0.5\columnwidth]{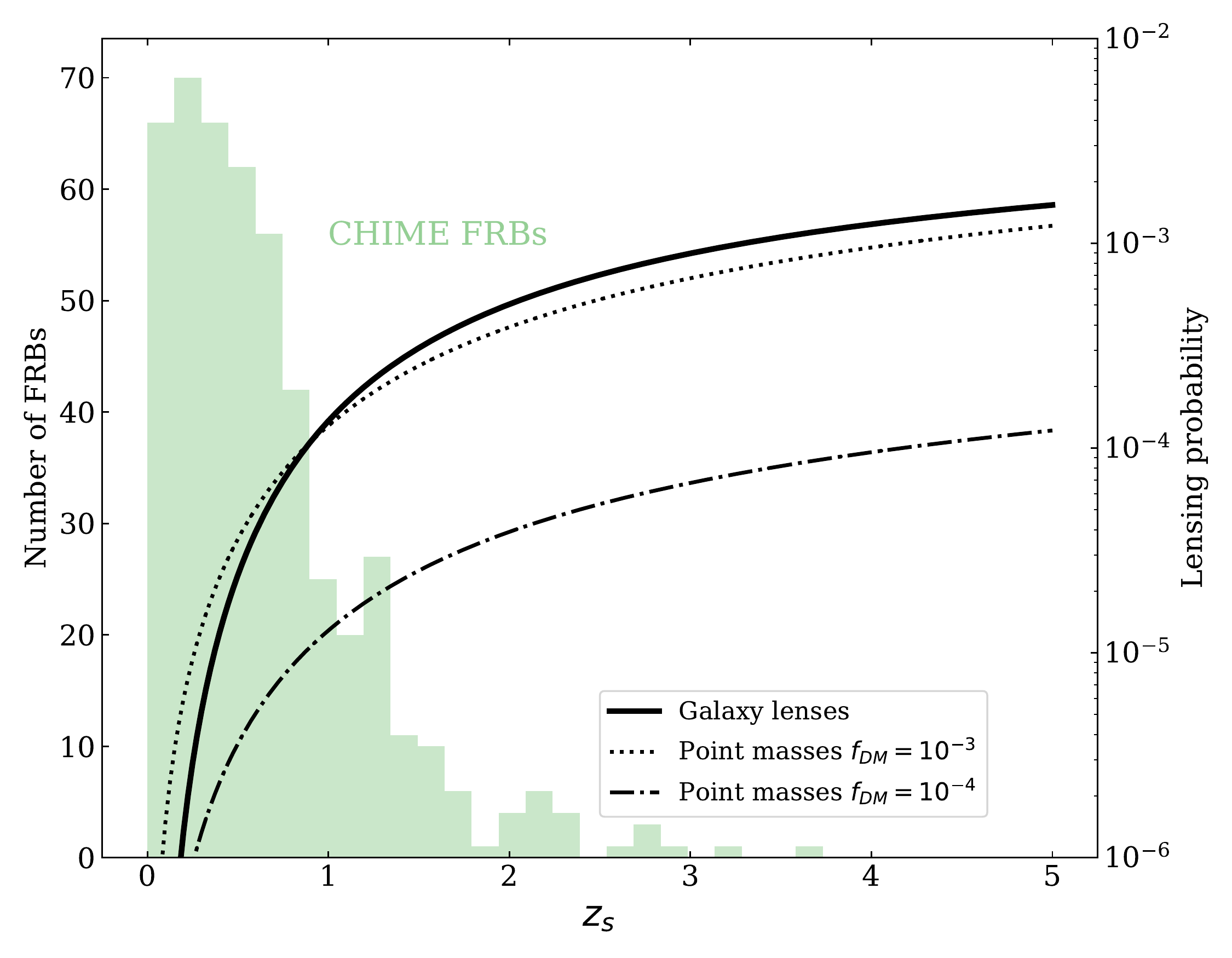}
    \caption{The modeled CHIME/FRB redshift distribution (green histogram, left vertical axis) and the probability that a source at redshift $z_s$ is 
    lensed (right vertical axis). 
    The lensing optical depth, 
    $\tau(z_s)$ increases quickly with redshift 
    such that most CHIME/FRBs have a very low 
    chance of strong lensing. We have assumed a  
    magnification bias $B=2$.}
    \label{fig:lens_prob}
\end{figure}

Clearly, the probability that a given 
FRB is lensed by an intervening galaxy is
strongly dependent on redshift, rising as 
$z^3$ for $z\lesssim1.5$. This is apparent in 
the solid black curve of Figure~\ref{fig:lens_prob}. Therefore, any 
FRB survey that hopes to detect gravitational lensing
would benefit from detecting distant bursts.

\subsection{Coherent gravitational lensing}
\label{coherent-lensing}
Radio telescopes measure directly the 
electric field of incoming electromagnetic waves, 
sampling voltages roughly one billion times per second 
for $\sim$\,decimeter wavelengths. 
FRB search pipelines ``detect'' this data by 
effectively squaring the voltages. They then downsample
in time and frequency to a manageable data rate, 
and search the lower-resolution intensity 
data for dispersed pulses. Many surveys now preserve 
the raw voltage data with a buffer that can be triggered and saved to disk, allowing astronomers 
to analyze the radio pulse's waveform. This is true 
for current surveys such as CHIME/FRB \citep{chime-baseband}, ASKAP \citep{bannister-2019}, DSA-110, and 
will be true for nearly all upcoming surveys.

Having access to the waveform itself is a major advantage 
for gravitational lensing \citep{Eichler2017}. 
If two pulses arrive from a 
similar sky position but at different times, their voltage 
timestream can be cross-correlated to test whether they 
are the same pulse (i.e. lensed copies of one another) 
and not just repeat bursts from the same source.
Crucially, this can be done on timescales shorter than the 
burst width using an autocorrelation. A ``single'' FRB can be correlated with itself, and one can search for power at non-zero time-lags, down 
to delays of the reciprocal bandwidth (nanoseconds). 
Similar techniques have been successfully applied to the voltage timestreams of 
giant pulses, effectively descattering 
Galactic pulsars \citep{main-2017}. Despite the 
theoretical limit of 1/$\Delta B$, for this paper we follow \citet{Wucknitz2021} and assume a practical time-lag (lensing delay) lower limit of 1\,microsecond. 

There are a number 
of practical issues related to coherent gravitational lensing 
searches. These include inverting the instrument's polyphase filterbank, removing dispersion measure (DM) with high precision, radio frequency interference (RFI), and the 
deleterious effects of interstellar scattering. However,
these are beyond the scope of our work and we point the 
reader to a detailed description and application 
of coherent lensing searches, or FRB gravitational 
lens interferometry, by the CHIME/FRB collaboration 
\citep{kader-2022, leung-2022}. 
We are here focused on detection rates 
and applications to cosmology and fundamental physics 
for upcoming surveys.

\section{FRB surveys}
\subsection{Detection rates}
\begin{figure}[htb]
    \centering
    \includegraphics[width=0.98\columnwidth]{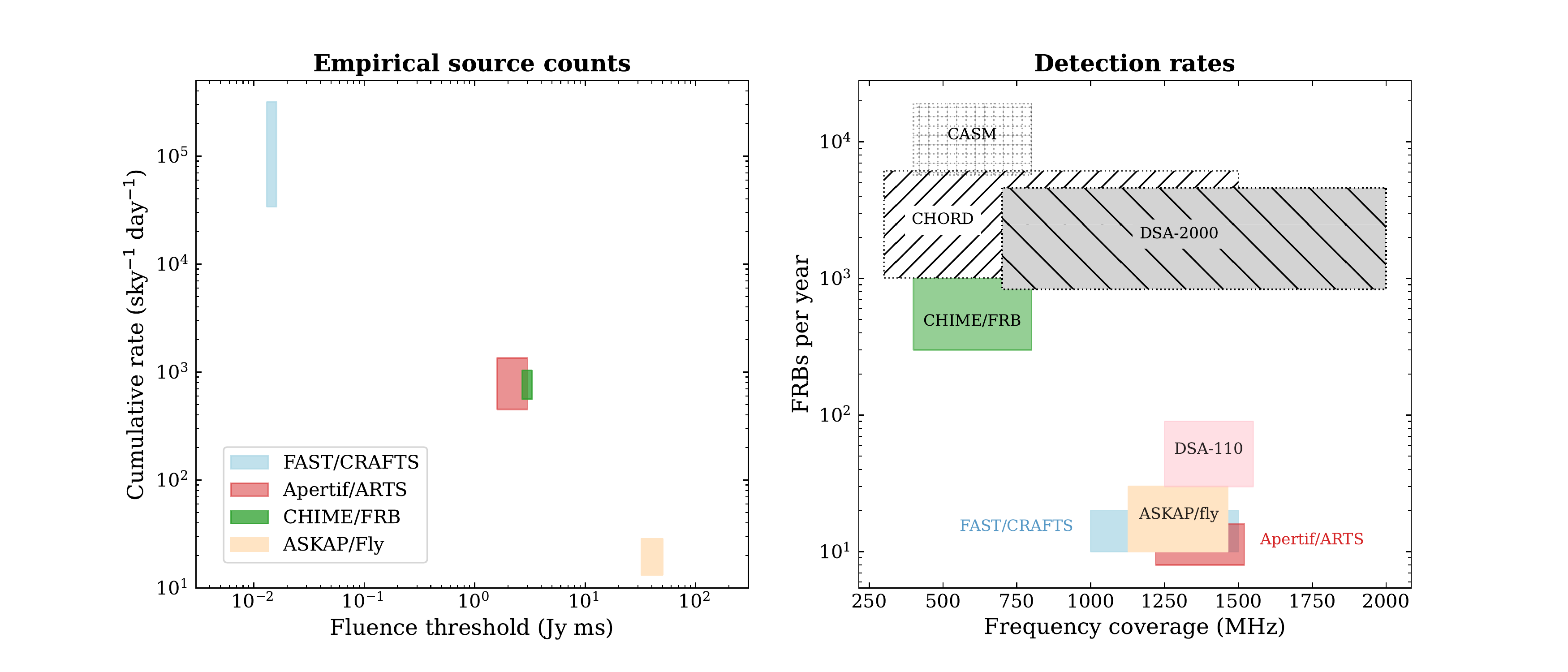}
    \caption{All sky FRB rates and survey detection rates. The left 
    panel shows the cumulative all-sky event rate of FRBs
    as a function of fluence threshold. The long baseline in 
    fluence provided by FAST and ASKAP in Fly's Eye mode allow 
    for source counts constraints over nearly four orders of magnitude. 
    The right panel shows \textit{effective} detection rates and 
    frequency coverage
    for four surveys currently on sky (solid boxes), as well as three 
    upcoming surveys for which we have run forecasts. The FRB detection rate
    here includes observing duty-cycle. For example Apertif only 
    observed $\sim$\,25$\%$ of the time so its effective detection rate 
    was one quarter of its rate per time on sky.}
    \label{fig:frbrates}
\end{figure}

Despite considerable advances in constraining  
the source counts and all-sky rate of FRBs, 
forecasting detection rates on new surveys remains 
challenging \citep{connor-2019}. This is due to disparate RFI environments and 
detection pipelines between telescopes as well as the 
unknown frequency dependence of the FRB rate. However, 
thanks to a large collection of FRBs from CHIME at 400--800\,MHz and $\mathcal{O}(100)$ events from surveys near 1.4\,GHz, we are no longer limited by small number statistics. 
These surveys have also implemented careful 
injection tests to measure pipeline completeness, thus we are in a far better position to estimate future survey detection rates than we have ever been in before. We 
also now have cumulative rate measurements spanning 
more than three orders of magnitude in flux density 
threshold, giving us a handle on the abundance of 
FRBs at 10\,mJy up to 100\,Jy \citep{fast-dm, shannon-nature, james-2019}. Figure~\ref{fig:frbrates} shows current FRB source counts 
and effective detection rates for several surveys. FAST has shown 
that the milliJansky radio sky is full of FRBs \citep{fast-dm}, which is
promising for sensitive future surveys such as DSA-2000 and CHORD.

Here we extrapolate from known surveys to estimate 
the detection rates at DSA-2000, CHORD, and the Omniscope. This 
will allow us to forecast gravitational lensing 
science that they can do. We start with the standard simplified rate equation 
for a transient survey, which is the product of 
the survey's FoV, $\Omega$ and areal source density above some minimum flux density, $n(>S_{min})$,

\begin{equation}
    \mathcal{R} = \Omega \, n(>S_{min}).
\end{equation}

Assume a survey $k$ has a system-equivalent flux 
density, $SEFD_k$, $n_{p,k}$ polarizations, 
and radio bandwidth, $B_k$. We can now extrapolate from 
a survey $l$ with a known 
detection rate of survey $\mathcal{R}_l$, giving,

\begin{equation}
    \mathcal{R}_k = \mathcal{R}_l \, \frac{\Omega_k}{\Omega_l} \left ( \frac{SEFD_l}{SEFD_k} \sqrt{\frac{B_k\,n_{p,k}}{B_l\,n_{p,l}}}\right )^\alpha.
\end{equation}

\noindent In Table~\ref{tab:surveys} we list 
the parameters for current and upcoming telescopes;
for existing surveys we have estimated the 
FRB detection rates. For parabolic reflectors, we assume $\Omega \approx 1.13\,\left (\frac{\lambda}{D} \right)^2$. 
Both DSA-2000 and CHORD 
utilize ultrawideband receivers, forcing us to make a 
choice about our treatment of the frequency dependence 
of FRB rates. Rather than modelling an FRB spectral index, 
we can break the surveys up into two sub-bands that are assumed 
to find different FRBs. This is partly because FRBs are often 
band-limited and because the FoV mismatch across an ultrawideband 
survey: For CHORD, the primary beam at the lowest frequencies 
is 25 times larger than at the top, so most FRBs arriving 
in the beam at the bottom of the band will not be detected 
at the top of the band, even if the pulses spanned $\sim$\,1.2\,GHz.
We combine empirical rates at 1.4\,GHz from 
FAST \citep{fast-dm} and Apertif \citep{Apertif-2022} with the CHIME/FRB detection rates at 600\,MHz.
We extrapolate directly from the CHIME rate (below 1\,GHz) and 
from the FAST/Apertif rate (above 1\,GHz, denoted by subscript ``$L$"). 
Breaking an ultra-wideband into two sub-bands, the rate equation becomes,

\begin{equation}
    \mathcal{R}_k = \mathcal{R}_{CH} \, \frac{\Omega_k}{\Omega_{CH}} \left ( \frac{SEFD_{CH}}{SEFD_k} \sqrt{\frac{B_k}{B_{CH}}}\right )^{\alpha}
    + \\ 
    \mathcal{R}_{L} \,  \frac{\Omega_k}{\Omega_{L}} \left ( \frac{SEFD_{L}}{SEFD_k} \sqrt{\frac{B_k}{B_{L}}}\right )^{\alpha}
    \label{eq:frb-rate}
\end{equation}

\begin{table*}
\center
\begin{tabular}{lccccccc}
\textbf{Survey}    & \textbf{FoV (deg$^2$)} & \textbf{SEFD (Jy)} & \textbf{Frequency (MHz)} & \multicolumn{1}{l}{$\mathbf{N}_{ant}$} & \multicolumn{1}{l}{\textbf{Diameter (m)}} & \multicolumn{1}{l}{\textbf{Rate (year$^{-1}$)}} & \multicolumn{1}{l}{\textbf{$\left < \mathbf{z_s} \right >$}} \\ \hline
CHIME              & 200                    & 50                 & 400--800                 & 1024                                    & N/A                                       & 500--1000           & 0.5                                 \\
Apertif            & 9                      & 75                 & 1220--1520               & 10                                      & 25                                        & 30--90              & 0.5                            \\ 
FAST            & 0.019                      & 1.33                 & 1000--1500               & 1                                      & 500/300$^\dagger$                                        & 4--40                  & 1.35                       \\ \hline
\textit{DSA-2000}  & 18--3                  & 2.5 / 6$^\dagger$                & 700--2000                & 2000                                    & 5                                         & 1000-4000                                  & 1.2             \\
\textit{CHORD}     & 72--2.9                & 9                  & 300--1500                & 512                                     & 6                                         & 1100-6000                       & 1.0                        \\
\textit{CASM} & 5000                   & 50                 & 400-800                  & 5000                                  & N/A                                       & 12500-25000    & 0.5                                   
\end{tabular}
\caption{The parameters for current and upcoming (italicized) FRB surveys. The rate column gives the FRB detections per one year on sky, and not the effective detection rate which requires multiplying by observing duty-cycle. The final 
column, \textbf{$\left < \mathbf{z_s} \right >$}, is the mean redshift. This is
estimated from the DMs of current surveys and from modelling for the three 
upcoming surveys (see Section~\ref{sec:redshifts}).
\\$\dagger$ corresponds to an 
effective value for the FRB search.}
\label{tab:surveys}
\end{table*}

\subsubsection{DSA-2000}
The DSA-2000 is a proposed wide-field radio camera 
that will have 2000$\times$5\,m steerable 
antennas and will observe between 700--2000\,MHz \citep{dsa-2000-whitepaper}.
It will have a system-equivalent flux density (SEFD) of 2.5\,Jy and 
a field-of-view (FoV) given by,

\begin{equation}
\Omega = 10.8\,\left(\frac{\nu}{1100\,\mathrm{MHz}}\right)^{-2}\,\,\mathrm{deg}^2.
\end{equation}

\noindent Utilizing the full FoV for pulsar and FRB search requires forming and searching a large 
number of beams, which is computationally challenging. 
This number is given by $N_{beam} = \left ( d_{max}/D\right)^2$, 
where $d_{max}\approx15.3$\,km is the longest baseline and 
$D=5$\,m is dish diameter, giving $N_{beam}\approx10^7$. For a
reduced number of beams there is a trade-off between 
the effective FoV that can be searched, $\Omega_{eff}$, and 
effective sensitivity. If baselines longer than $d_{cut}$ 
are discarded, the effective SEFD will be 2.5\,Jy\,$\sqrt{f(d<d_{cut})}^{-1}$, where $f(d<d_{cut})$ 
is the fraction of baselines shorter than $d_{cut}$. The effective FoV at 1100\,MHz will then be,

\begin{equation}
    \Omega_{eff} = min\left \{ 10.6\,\mathrm{deg}^2, N_{beam}\times1.13\,\left(\frac{c/1100\,\mathrm{MHz}}{\,d_{cut}}\,180 / \pi\right)^2 \right \},
\end{equation}

\noindent where $N_{beam}$ is the total number of beams that can be searched. We use a $min$ operator because we cannot usefully search 
beyond the instrument's primary beam. Assuming Euclidean source 
counts, detection rate is maximized when $d_{cut}$ is  
such that the full primary beam is tiled by $N_{beam}$.
We find that if DSA-2000 can afford to search $10^6$
beams, $d_{cut}$ should be chosen such that the synthesized beamwidth 
is 11'' and the effective SEFD is 5\,Jy. For the current antenna 
configuration, this corresponds to $d_{cut}\approx4.1$\,km. 
For the remaining forecasts, we assume that DSA-2000 will have 
SEFD$=$6\,Jy and $\Omega_{eff}=10.6$\,deg$^2$ at 1100\,MHz. We 
further assume that DSA-2000 will only search for 
FRBs between 700\,MHz and 1600\,MHz.

Following Equation~\ref{eq:frb-rate}, we find that DSA-2000 
will detect 830--4600 FRBs per year. Over the course of its nominal 
five year survey (and an estimated four years on sky), this would result 
in $\mathcal{O}(10^4)$ localized FRBs in which to search for 
gravitational lensing.

\subsubsection{CHORD}
CHORD is funded transit radio telescope that is preparing for construction \citep{Vanderlinde-2019}.
The FRB search on CHORD will use 512$\times$6\,m antennas 
organized in a compact grid, comprising the array's core. 
CHORD employs ultra-wideband antennas covering 300--1500\,MHz 
and has an expected SEFD of 9\,Jy. Its FoV is given by, 

\begin{equation}
\Omega = 7.7\,\left(\frac{\nu}{1100\,\mathrm{MHz}}\right)^{-2}\,\,\mathrm{deg}^2.
\end{equation}

Though it has less 
collecting area and sensitivity than DSA-2000, the dense 
antenna configuration of its core is optimal for searching the 
full FoV at nominal sensitivity. It will only need to 
form and search 512 beams. Two outrigger stations separated 
from the core by more than $10^3$\,km allow for VLBI localization 
of FRBs at milli-arcsecond-level precision. CHORD has an unprecedented 
5:1 radio bandwidth that will require sub-band searching, due to the FoV 
mismatch as a function of frequency \citep{Vanderlinde-2019}. 
Rather than 
break the full band into just two sub-bands as we have done with DSA-2000, 
we choose to sum FRB detections in three geometric sub-bands: 300--515\,MHz,
515--880\,MHz, and 880--1500\,MHz. In the top sub-band, between 880--1500\,MHz, 
we extrapolate from the FAST detection rate because of the match between sensitivity and frequency between the two 
telescopes. In the bottom two sub-bands, 
we extrapolate from CHIME/FRB and use $\alpha=1.5$. We 
find that CHORD will discover 1000--6000 FRBs per year, 
all with voltage dumps and precise localizations.  

\subsubsection{A coherent all-sky monitor (CASM)}
The logical endpoint of the 
large-$N$ small-$D$ paradigm for 
radio interferometers is an aperture array, in which radio receivers are pointed directly at the sky without a light-focusing dish. A dense aperture array with a large 
number of feeds could survey thousands of square degrees 
simultaneously at high sensitivity, detecting an enormous number of FRBs. Such an instrument would be a coherent 
all-sky monitor, so we refer to this hypothetical 
survey throughout the remainder 
of the paper with the acronym ``CASM''. 

Front-end electronics and digitization would be challenging for such a large number of antennas, but 
beamforming would be made feasible by FFT beamforming \citep{peterson-fft, tegmark-fft}. 
For example, a feed with a $\sim$\,70\,deg opening angle would have a FoV 
that is $\sim$\,10$^4$ times larger than that of Parkes and 25 times greater than CHIME. 
Each feed would have an effective collecting 
area of $\lambda^2$. To build up the same collecting area 
as CHIME at 600\,MHz, roughly 25,000 feeds would be required.

Dense aperture arrays already exist, but not 
with the ability to use the full FoV coherently. 
For example, 
the Electronic Multi Beam Radio Astronomy ConcEpt (EMBRACE)
was designed and constructed in the Netherlands in order to 
demonstrate phased array technology for the Square Kilometer Array (SKA) \citep{embrace}. EMBRACE has over 20,000 elements and roughly 160\,m$^2$ of collecting area. It uses hierarchical analog beamforming. 
The dense aperture array design once proposed for the 
SKA, and for which EMBRACE was a pathfinder, will not be built. This has led some work to mistakenly infer that SKA-Mid will detect as many as 10$^7$ 
FRBs per year \citep{Hashimoto-ska, chen-II}, based on the dense aperture array proposal.

There is at least one proposed FRB survey that falls within the dense aperture array design. The Bustling Universe Radio Survey Telescope in Taiwan (BURSTT) is an example of 
a coherent all-sky monitor that would have a high 
FRB detection rate and outriggers for localization \citep{burstt-2022}. 
It has been considered explicitly in the 
context of coherent gravitational lensing. 
Beyond the high FRB detection rate afforded by an ultra-widefield telescope, 
CASM experiments can go after 
longer time-delays from lensing because a large patch of sky can 
be observed continuously. Outrigger stations will provide 
VLBI localizations that will enable new applications of 
FRB gravitational lensing by massive galaxies. BURSTT-2048 is expected 
to have an SEFD of 600\,Jy and will spend a significant portion of its 
time near the north celestial pole \citep{burstt-2022}.

For the purposes of this paper, we will assume a future 
CASM survey that has the sensitivity of CHIME/FRB. 
The SEFD is taken to be 50\,Jy, observing in the band 400--800\,MHz, 
with a FoV of 5,000\,deg$^2$. In other words, a CHIME/FRB-like survey
but with 25\,$\times$ the sky coverage. These survey 
parameters make forecasting FRB detection rates very simple, 
because it will be roughly 25 times the CHIME/FRB detection rate; 
one does not have to assume anything about spectral index
nor source counts. The major uncertainty, instead, 
is whether such an interferometer can be built to spec. 
We do not attempt to model system performance uncertainty, 
and instead use CASM as a place holder for an all-sky survey 
with CHIME/FRB sensitivity observing at 400--800\,MHz. 
The inferred detection rate of CASM would be 7,500--25,000 
FRBs per year, reaching $\mathcal{O}(10^5)$ FRBs after 
5--10 years on sky.

\subsection{Redshift distribution}
\label{sec:redshifts}

\begin{figure}[htb]
    \centering
    \includegraphics[width=0.7\columnwidth]{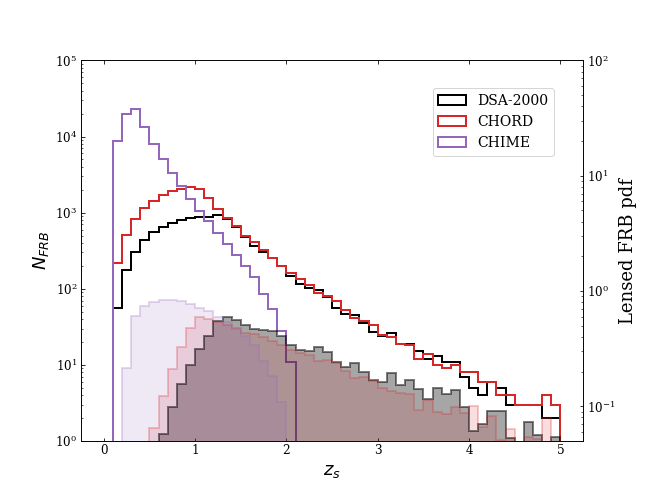}
    \caption{The modeled redshift distributions of DSA-2000,
    CHORD, and CASM. The latter has the highest total 
    detection rate, but DSA-2000 and CHORD will probe higher 
    redshifts thanks to their high sensitivity. The shaded histograms represent each survey's lensing probability density function.}
    \label{fig:frbredshift}
\end{figure}

The number of lensed FRBs in upcoming surveys will be a strong 
function of the source redshift distribution: lensing optical 
depth increases as the cube of redshift for $z_s\lesssim1.5$ 
for strong galaxy lensing. 
It is imperative that a forecast for the lensing 
detection rate includes a realistic model for the 
FRB redshift distribution. To address this, we make use of 
two important relationships. One is the dependence of 
observed extragalactic dispersion measure (DM$_{ex}$) on redshift, and the other is the 
relationship between distance and burst brightness. 

\citet{macquart-dm}
established that FRBs with higher DM$_{ex}$ 
are typically farther away, as expected if the IGM dominates dispersion. The 
so-called ``Macquart relation'' is given by the 
following,

\begin{equation}
\left < \mathrm{DM}_{IGM}(z) \right > \approx 865\,z\,\,\,\mathrm{pc}\,\mathrm{cm}^{-3}.
\end{equation}

\noindent While this approximate linear relationship holds, we caution that the $\mathrm{DM}_{ex}$ dependence on $z$ is impacted by the host galaxy DM distribution \citep{james-2022}. We solve 
for $z$ with,

\begin{equation}
\mathrm{DM}_{ex} = 865\,z + \frac{100}{1+z}\,\,\,\,\,\,\mathrm{pc}\,\mathrm{cm}^{-3},
\end{equation}

\noindent which can be rearranged into a 
quadratic equation,

\begin{equation}
865\,z^2 + (\mathrm{DM}_{ex} - 865)z  + \mathrm{DM}_{ex} - 100 = 0.
\label{eq:dmz}
\end{equation}

The CHIME/FRB Catalog 1 \citep{chime-catalog1} estimates extragalactic DM 
by subtracting the expected contribution of the Milky 
Way along that line of sight from the observed DM. 
We use these to solve for $z$ in Equation~\ref{eq:dmz}.
The resulting redshift distribution for the first 
CHIME/FRB release is shown in Figure~\ref{fig:lens_prob}

The relationship between FRB brightness and
distance is critical for predicting the redshift 
distribution for more sensitive surveys. Fortunately 
for telescopes such as DSA-2000, CHORD, and FAST, 
there is evidence that dimmer FRBs come from higher 
redshifts, meaning more sensitive surveys will 
probe a deeper redshift distribution. This was not guaranteed, as a sufficiently flat luminosity function 
would result in the brightest FRBs being farther away
\citep{macquart-ekers}. The positive correlation between 
fluence and DM was 
demonstrated by \citet{shannon-nature} when comparing 
the Parkes DM$_{ex}$ distribution with that of 
ASKAP in Fly's eye mode. The relationship has 
also been borne out by 
the large DMs of FAST-discovered FRBs, 
which can detect pulses down to tens of milliJanskys \citep{fast-dm}.
The exact mapping between observed fluence and redshift 
will depend on the FRB luminosity function, 
the \textit{true} source distribution in $z$, and selection effects \citep{connor-2019}, but for our purpose this simple model 
is adequate. We take the DM and redshift distribution 
of the hypothetical CASM survey to be the same as CHIME/FRB, since 
it will have similar parameters to CHIME
but with 25 times more FoV.

In a Euclidean volume, the mean FRB distance in a given 
survey scales as the square-root of a survey's sensitivity \citep{dongzi-2019}, so cutting in half 
a telescope's SEFD will lead to it detecting 
FRBs that are, 
one average, $\sim$\,40\,$\%$ farther away. For a 
cosmological population, the relationship is 
slightly weaker. We simulate $n(z)$ 
for DSA-2000 and CHORD based on the modeled 
redshift distribution of CHIME/FRB. 
We assume FRBs have a constant comoving volume density and 
a single cumulative power-law luminosity function that has 
$\gamma=1$.

\section{Results}
\label{section:results}

\begin{table}
\begin{center}
\begin{tabular}{lcccc}
\textbf{}   & \textbf{DSA-2000} & \textbf{CHORD} & \textbf{CASM} & \textbf{CHIME/FRB} \\ \hline
$N_{FRB}$ (5 years) &  4,000--16,000  &        4,400--24,000   &  50,000--100,000   &  2,500--5,000    \\
$N_{lens, *}$   &      6--32              &        8--48        &  50--100   &  0.5--5          \\ 
$f_{PBH}$ (0.1--100\,$M_\odot$)$^\dagger$              &      $\leq 1.2\times10^{-3}$               &        $\leq 1.1\times10^{-3}$          &  $\leq 1.0\times10^{-3}$  &  $\leq 0.06$          \\
$\Omega_{IMBH}$ ($10^3$--$10^5$ $M_\odot$)$^{*}$       &      $\leq 10^{-2}$               &        $\leq 10^{-2}$          &  $\leq 10^{-3}$  &  $\leq 10^{-1}$              \\
$N_{lens, IMBH}$ ($10^3$--$10^5$ $M_\odot$)$^*$       &      0.1--7.3               &        0.1--7.0          & 0.3--7.2  &  0.01--0.29              \\

$N_{lens, gal}$ ($10^{10}$--$10^{12}$ $M_\odot$)$^{**}$            & -                      & -                 & 5--40 & -  \\ \hline
\end{tabular}
\caption{The results of our FRB gravitational lensing forecasts 
for four surveys, assuming five years of operation 
with $\sim$\,80$\%$ duty-cycle. 
\\$\dagger$ The constraints on the fraction of matter in 
primordial black holes without a detection of FRB microlensing.
\\$*$ We take $\Omega_{IMBH}$ to be the constraints on the density of 
intermediate mass black holes in the absence of a lensing detection. 
$N_{lens, IMBH}$
corresponds to the expected number of FRB lensing detections with $\Delta t$ between 
1--100 seconds assuming the rate from BATSE GRBs \citep{Paynter-2022}.
\\$**$ The expected number of detections of FRBs lensed by massive galaxies. DSA-2000, 
CHORD, and CHIME/FRB cannot access time delays longer than a pointing duration/transit time, 
so they will not see strong lensing by foreground galaxies. The CHIME/FRB stellar lensing uncertainty is large because it is not clear what fraction of total events will have 
preserved voltage data and high DM.} 
\end{center}
\label{tab:results}
\end{table}

\subsection{Strong lensing by stars}
Considerable attention has been paid to short-duration FRB gravitational 
lensing by MACHOs and PBHs \citep{munos2016, zhou-PBH, kader-2022, leung-2022}. To our knowledge, nobody has 
considered strong lensing of FRBs by stars in the cosmological 
context. Stars at $z_l=0.5$ will have an Einstein radius 
of $\theta_E \approx 1.6''\times\left ( \frac{M}{M_\odot} \right )^{1/2}$ for a source at $z_s=1$.
This corresponds to a physical impact parameter 
of roughly $10^{16}$\,cm, much larger 
than stellar radii. The geometry gives a fiducial 
lensing timescale of 20\,microseconds per solar mass, 
which is already within the region of lags that 
CHIME/FRB is capable of searching \citep{kader-2022, leung-2022}. If the cosmological stellar density parameter is 
$\Omega_*\approx 0.0025$, then this is equivalent to 
$f_{*} = \frac{\Omega_*}{\Omega_{c}} \approx 10^{-2}$
\citep{fuk, planck18}. From Figure~\ref{fig:lens_prob}, we see that the optical depth 
for such a population of point-masses is significant, even for a relatively nearby CHIME/FRB catalog. 

However, unlike PBHs and MACHOs, stars are 
highly concentrated near the cores of galaxies. \citet{wyithe-turner} have shown that 
the strong spatial clustering of stars renders 
the Press-Gunn approximation inadequate for 
cosmological stellar microlensing. In other words, 
if the isolated lens assumptions is broken and $\tau$ 
approaches or exceeds 1, point-mass lenses may be 
over-counted \citep{koopmans}.
An extreme example is if stellar lenses in the Universe 
were all aligned radially; the probability of strong 
microlensing would be effectively zero.
\citet{wyithe-turner} found that roughly 1$\%$ of sources at or beyond redshift 2 will be lensed by stars. Their work, and related papers,
were in the context of GRBs, where multiple 
copies from stellar microlensing cannot be detected unless the delay is 
longer than the burst width. FRBs do not have this issue.
Here we consider how 
stars in elliptical and spiral galaxies will impact 
FRB gravitational lensing for realistic FRB redshifts, correcting for the spatial distribution of stellar mass in galaxies. In Section~\ref{sec:pbh}, 
we discuss how lensing by stars can be distinguished 
from other point-masses such as PBHs. 

For our forecast of FRB stellar lensing, we make 
the conservative assumption that only stars 
within a relatively small threshold impact parameter of a 
foreground galaxy will contribute to the optical 
depth. The probability of stellar lensing 
is roughly 1 for sources that pass within 
$\sim$\,3\,kpc of a Milky Way like galaxy. 
Such galaxies have a volumetric density of 0.01\,Mpc$^{-3}$,
which means the optical depth for stellar lensing would be $\tau_* \approx 3\times10^{-4}\,\left ( \frac{D_l}{1\,\mathrm{Gpc}} \right )$ for spiral galaxies. For elliptical 
galaxies the 
impact parameter within which microlensing optical 
depth is $\sim$\,1, is larger. Assuming sightlines 
within 6\,kpc of elliptical galaxies are microlensed, we find $\tau_* \approx 10^{-3}\,\left ( \frac{D_l}{1\,\mathrm{Gpc}} \right )$. This leads to the striking fact that roughly one 
per thousand CHIME/FRB sources could be lensed 
by stars on timescales between 1--100\,microseconds. 

Similar to strong lensing by galaxies, the majority of
stellar microlensing events will be due to elliptical galaxies. \citet{wyithe-turner} find that for spiral galaxies, roughly 70$\%$ of high-$z$ sources microlensed by stars will also be strongly lensed by the galaxy itself. For elliptical/S0 galaxies, this number 
is closer to half. Therefore, detecting an 
FRB with microsecond lensing is a good indicator that 
it is also lensed on days to weeks timescales 
and that its host galaxy will be lensed in 
an optical image. If the FRB source is a repeater, 
then the lensing system could be monitored in 
order to constrain $H_0$ \citep{Wucknitz2021}.

Current and future wide-area optical/IR surveys will prove excellent resources for identifying the host galaxies of stellar microlenses. The extant Legacy Survey delivers images over 14,000\,deg$^{2}$ with a 1.2\arcsec~point-spread function (PSF) FWHM in the $g$, $r$, and $z$ bands, detecting objects to $g=24.0$ and $r=23.4$ \citep{legacy}. Space-based imaging surveys with \textit{Euclid} \citep{euclid} will span 15,000\,deg$^{2}$ across several visible and near-IR bands, with detection limits of 26.2\,mag in the visible and 24.5 in the near-IR, and a PSF FWHM of 0.225\arcsec~in the visible and 0.3\arcsec~pixels in the near-IR that undersample the PSF. The Nancy Grace Roman Space Telescope \citep[Roman;][]{roman} will cover only 2,000\,deg$^{2}$ in the near-IR with a 0.28\arcsec~PSF FWHM, but will detect objects to 26.6\,mag. Finally, the SPHEREx mission will deliver accurate redshifts of a few $\times10^{8}$ galaxies over the whole sky, with a detection limit of 18.4\,mag in the near-IR in each spectral channel. 

\begin{figure}[htb]
    \centering
    \includegraphics[width=0.65\columnwidth]{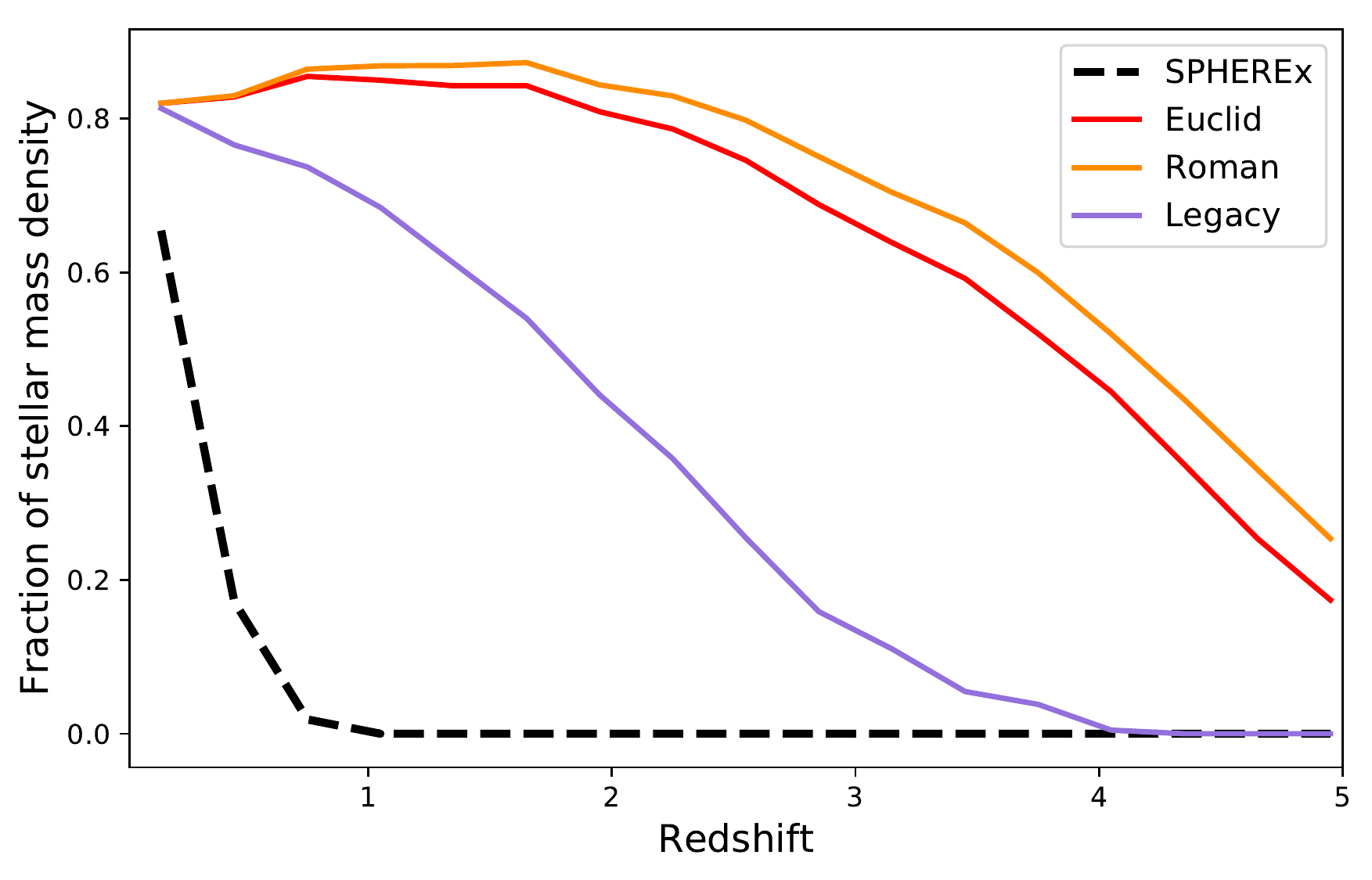}
    \caption{The fraction of the cosmic stellar mass density recovered within the cones of the Legacy Survey (solid purple), the \textit{Euclid} (solid orange) and Roman (solid red) wide-area surveys, and SPHEREx (dashed black). For SPHEREx we assume that objects must be detected in at least one spectral channel, and for the remainder of the surveys we assume that objects must be detected in at least one imaging band. See text for further details on the surveys, and the simulated galaxy catalog used to derive the results.}
    \label{fig:stellarmass_detect}
\end{figure}

Figure~\ref{fig:stellarmass_detect} shows the fraction of the stellar mass density of the Universe recovered within each survey footprint, as a function of redshift. We used the survey detection thresholds to identify detectable galaxies at each redshift in the output catalog of a recent semi-analytic galaxy formation model based on the Millennium simulation rescaled to the latest Planck cosmology \citep{mpasim}. We used predicted galaxy magnitudes that included the simulated effects of internal dust extinction, and required detection in at least one band for each survey (at least one spectral channel for SPHEREx). These results suggest that a significant fraction of the host galaxies of stellar microlenses will be identified within the \textit{Euclid} and Roman survey areas, potentially enabling modeling of the galaxy-scale lens mass distribution using the lensed FRB host galaxy, as well as studies of the distant FRB host galaxy itself. The former can be used to predict the potential observation of future images of the FRB lensed on the scale of the galaxy \citep[e.g.,][]{Rodney-2021}. 

Astronomers must note the risk 
of mis-identifying a lens galaxy as the FRB host. Lensing searches should be done 
on microsecond timescales to differentiate 
the host from the lens, especially if the 
extragalactic DM is larger than expected. 
This will affect roughly $10^{-3}$ 
sight lines for typical FRB distances. Since most lenses will be elliptical galaxies but most FRB hosts appear to 
be star-forming, a tight association with an elliptical galaxy may suggest lensing. Such localization precision can currently be achieved for CHIME/FRB VLBI, DSA-110, ASKAP, and Meerkat\citep{meerkat}. 
Another practical issue is that scattering in the lens galaxy may diminish the signal's significance, particularly in the case of spiral galaxies, but less so with ellipticals. 



\subsection{Primordial black holes}
\label{sec:pbh}
\begin{figure}[htb]
    \centering
    \includegraphics[width=0.7\columnwidth]{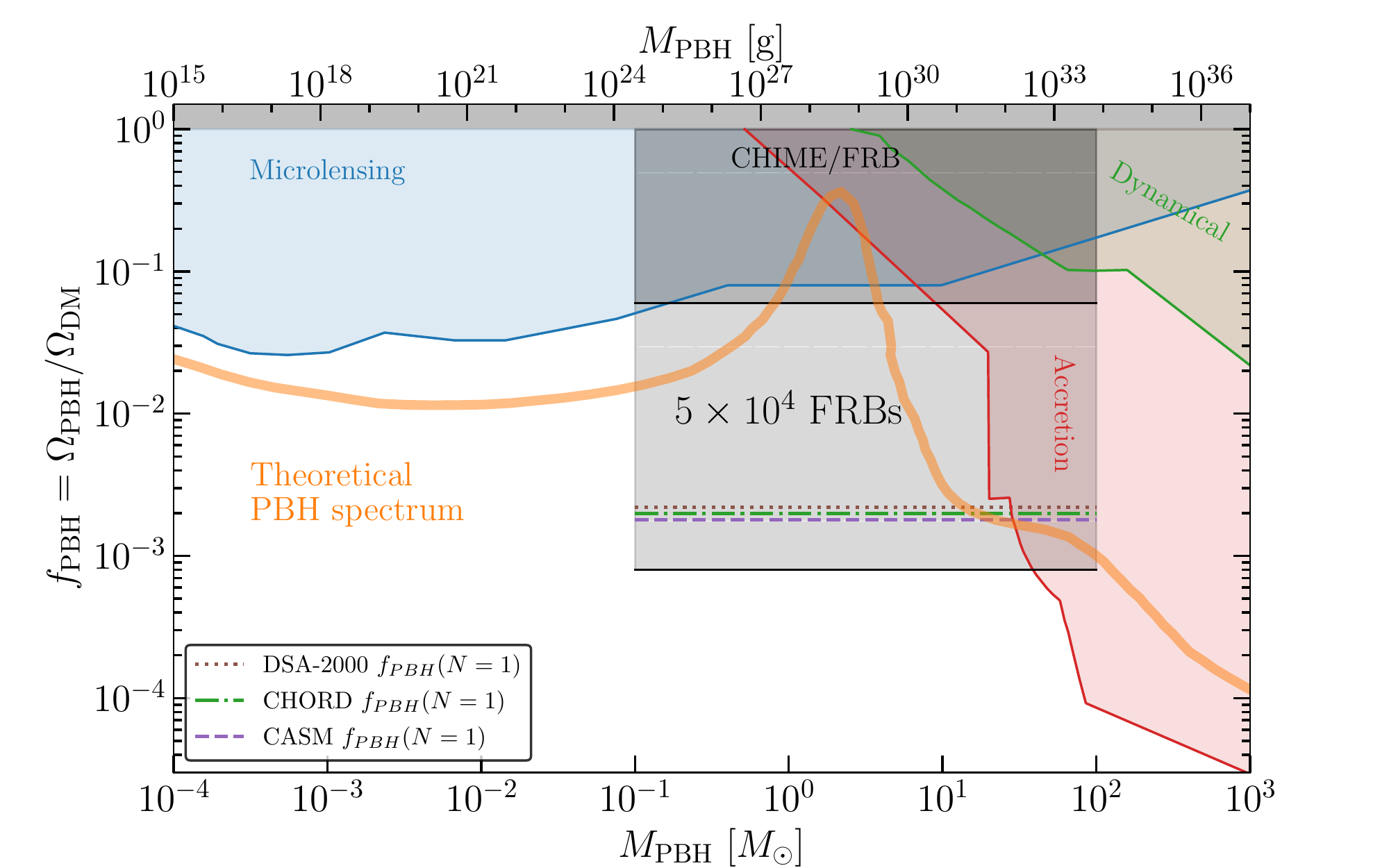}
    \caption{The current constraints on the mass of 
    primordial black holes along with future constraints from 
    FRBs. If no FRB is lensed on timescales 1\,$\mu$s to 
    10\,ms, $5\times10^4$ sources could constrain 
    $f_{PBH}$ to less than $8\times10^{-4}$ between 0.1--100\,$M_\odot$. 
    If CHIME/FRB 
    can coherently search for lensing in $\sim$\,500 
    unscattered FRBs, 
    a non-detection will rule out the region 
    $f_{PBH}/M_{PBH}$ space that is grey and double hatched. 
    The three horizontal lines correspond to the value of $f_{PBH}$ 
    that would produce on average 1 microlensing event per 
    year from PBHs for three upcoming FRB surveys. We created 
this figure by modifying existing plotting code https://github.com/bradkav/PBHbounds.}
    \label{fig:pbh}
\end{figure}

Primordial black holes (PBHs) are a theoretical class 
of black holes that form in the early Universe.
They may form via direct collapse from primordial fluctuations, or though other mechanisms \citep{pbh-review2020}. 
Unlike black holes that are formed after stellar collapse 
above the Tolman–Oppenheimer–Volkoff limit of $\sim$\,3\,M$_\odot$, 
PBHs can be produced with a wide range of masses. Constraining 
this mass spectrum is an active area of research and uses 
a variety of inputs, including evaporation timescale, 
small-scale CMB fluctuations, and Ly$\alpha$ forest \citep{pbh-review2020}. For simplicity, 
we consider here only a monochromatic mass distribution (MMD) 
such that all PBHs have the same mass, $M_{\rm PBH}$. 

We take $f_{\mathrm{PBH}}\equiv\frac{\Omega_{\mathrm{PBH}}}{\Omega_{c}}$ 
to be the fraction of cold dark matter that resides in primordial black holes, and estimate the region of $f_{\mathrm{PBH}}/M_{\mathrm{PBH}}$ parameter space 
that current and upcoming FRB surveys will be able to search. 
In Figure~\ref{fig:pbh} we show constraints on the abundance of primordial 
black holes assuming each survey 
will be able to detect lensing events with 
delays from microseconds to milliseconds ($M_L\sim\,0.1-100\,M_\odot$).

We find that a non-detection from 5 years on sky 
with DSA-2000, CHORD, and CASM produce a similar 
upper-limit of $f_{\mathrm{PBH}} \lesssim 10^{-3}$. 
While the CASM considered 
here would find many more FRBs, the deeper 
redshift distribution of DSA-2000 and CHORD 
result in a similar mean lensing optical depth. 
We plot the value of 
$f_{\mathrm{PBH}}$ that would produce an average of 1 lensed FRB per year 
for DSA-2000 (brown, dotted), CHORD (green, dot-dashed), and CASM (purple, dashed).

In \citet{zhou-PBH}, the authors calculate the $f_{\mathrm{PBH}}$ constraints for the non-detection of gravitational lensing 
in 593 CHIME/FRBs. However, their limits are weakened by the 
(then valid) assumption of incoherent lensing searches, where 
the minimum lensing delay is set by the CHIME FRB pulse width. 
They could not constrain PBH masses that are 
less than roughly 10\,$M_\odot$. The CHIME/FRB collaboration 
has now produced the first observational constraints 
from \textit{coherent} gravitational lensing \citep{kader-2022, leung-2022}. They have used the voltage data for 172 bursts from 114 independent sightlines to constrain the fraction of dark matter in PBHs with 
masses between $10^{-4}$ and 10$^4$ M$_\odot$. They consider 
the impacts of scattering from plasma local to the source and conservatively assume that high DM events have significant 
dispersion in the host galaxy, putting DM\,$>$\,500\,pc\,cm$^{-3}$
at smaller $z_s$. Due to memory constraints 
in their data buffer, voltage data for sources with DM\,$>$\,1000\,pc\,cm$^{-3}$ are not included in the sample \citep{leung-2022}. The authors place an upper limit 
of 0.8 on $f_{\mathrm{PBH}}$ for $M_{\mathrm{PBH}}\sim10^{-3}$\,M$_\odot$. 

CHIME/FRB has now detected over 3000 FRBs, so just $\sim$\,3$\%$ have gone into constraining $f_{\mathrm{PBH}}$
\citep{leung-2022}. 
If CHIME/FRB can search 
coherently for lensing events using the preserved voltage 
data of $\sim$\,500 FRBs, the PBH 
constraints will be given by the 
double-hatched grey region shown in Figure~\ref{fig:pbh}. We have assumed 
that FRBs with DM$>$1000\,pc\,cm$^{-3}$ or $z_s\gtrsim0.8$ do 
not contribute to the optical depth, because CHIME/FRB cannot store their voltages. 
For 50,000 FRBs with the same redshift distribution as CHIME/FRB, 
the grey singly-hatched block shows the region of $f_{\mathrm{PBH}}/M_{\mathrm{PBH}}$ parameter space that will have been ruled out (i.e. several years of observing with CASM). Note our upper-limits are less 
conservative than those of the CHIME/FRB collaboration 
because we have not included instrumental and 
propagation effects \citep{kader-2022, leung-2022}.

There is a further practical problem associated with searching for 
point-mass lensing events. Lensing by stars will likely dominate 
in the microseconds lag range, as 
explained in the previous section. Therefore, with only a 
time-delay measurement, it will be difficult to know if a 
PBH has been detected rather than a stellar lensing event 
in an intervening galaxy. This degeneracy can be mollified---if not fully eliminated---by 
noting 
that PBHs ought to be less concentrated than stellar mass, and they will 
be more uniformly distributed in the cold dark matter halos of galaxies. Thus, a strong microlensing event that occurs within 
$\sim$\,10\,kpc of the host galaxy center will likely 
be due to a star. Constraints on $f_{PBH}$ could  
be limited to sources with larger impact parameters. 
Using the spatial offset information requires that 
the telescope has $5''$ localization precision or better. At 
present, CHIME/FRB cannot achieve such localizations, so if 
there were a positive detection between 1--100\,$\mu$s, 
they could not attribute it to a PBH rather  
than a star. 

\subsection{Intermediate mass black holes}

IMBHs are black holes whose mass lies between 
stellar mass black holes and 
the supermassive black holes that reside 
at the centers of galaxies \citep{greene-review}. They are 
often defined to have masses between 
100\,M$_{\odot}$ and 10$^5$\,M$_{\odot}$. 
There is currently no concrete 
observational evidence for their existence, 
though there are a number of candidates 
in that range (e.g. ultra-luminous X-ray sources \citep{ulx-2017}). Theoretically, IMBHs ought 
to exist in relatively high abundance, as 
there must be an evolutionary bridge between stellar and supermassive black holes. 
Directly observing IMBHs and constraining 
their number density is major outstanding 
problem in astronomy. 

Lensed FRBs will probe the 
volumetric density of IMBHs. For a 
lens at redshift 0.5, the lensing time delay 
is between 3\,ms and a few seconds for lens masses of 10$^2$--10$^5$\,M$_\odot$. We parameterize 
the cosmological density of IMBHs in the standard way as,

\begin{equation}
    \Omega_{IMBH} = \frac{\rho_{IMBH}}{\rho_c},
\end{equation}

\noindent where $\rho_{IMBH}$ is the mean density of 
IMBHs and $\rho_c$ the critical density of the Universe,

\begin{equation}
\rho_c = \frac{3 H^2_0}{8\pi\,G}. 
\end{equation}

\noindent To forecast ``millilensing'' events 
on upcoming surveys, we consider two scenarios. First, we apply a
similar treatment of PBHs, where we ask how well 
$\Omega_{IMBH}$ can be constrained if no FRBs are lensed on 
timescales of 3\,ms to a few seconds. Our second approach 
is to assume previous candidates in the relevant 
mass range \citep{Paynter-2022, Vedantham-2017} were true gravitational lensing events and 
then extrapolate from their inferred optical depths.

If no millilensing event is found, we can constrain 
the cosmic density of IMBHs at $90\%$ confidence to,


\begin{equation}
     \tau \lesssim 2.3\, B^{-1}\,N_{FRB}^{-1}.
\end{equation}

\noindent The factor of 2.3 is from the 90$\%$ Poissonian 
confidence limits, having seen zero events. $B$ is magnification bias. We 
can then use the Press-Gunn relation to get,

\begin{equation}
    \Omega_{\mathrm{IMBH}}\lesssim 2.3\, B^{-1}\,N_{FRB}^{-1} \left ( \frac{\left <z \right >}{2} \right )^2
\end{equation}

We use the modeled redshift distribution means of DSA-2000, CHORD, 
CASM, and CHIME/FRB to compute these upper limits. They 
are displayed in Table~\ref{tab:results}.

Next, we will assume that gravitationally lensed fast transients 
have already been observed and extrapolate rates directly from them. \citet{Paynter-2022} found evidence that GRB\,950830 
was lensed by a $\sim$\,$5\times10^4$\,M$_\odot$ black hole 
at $z\approx1$. They also use the approximation \citep{press73} that 
$\Omega_{IMBH}\sim\tau(\left < z_s \right > = 2)$ and calculate 
optical depth from one lensed event in $\sim$\,2700 BATSE 
GRBs. For a mean source redshift of $\sim$\,2 they claim,

\begin{equation}
    \Omega_{\mathrm{IMBH}}(M\sim10^{4-5}\,\mathrm{M}_\odot)\approx4.6^{+9.8}_{-3.3}\times10^{-4}.
\end{equation}

\noindent A related result comes from ``Symmetric Achromatic Variability'' seen in active galaxies. \citet{Vedantham-2017}
proposed that the achromatic temporal variation observed 
in BL Lac object J1415+1320 was a millilensing event, plausibly caused by a dark matter subhalo or black hole located
within an intervening galaxy. The relatively large density of IMBHs implied 
by GRB\,950830 and J1415+1320 would be promising for upcoming FRB surveys, which 
will be sensitive to a wider range of lens masses and 
will have a larger collection of transients than BATSE. 
FRB lensing will be less ambiguous than GRB events, because of the coherent temporal 
methods described in Section~\ref{coherent-lensing}. 
Again, we use the 
modeled mean redshifts of four surveys
in Table~\ref{tab:surveys} and the $\tau(z_s)$ relation
to estimate the optical depth 
from IMBHs for FRBs, scaling from the mean redshift of  GRBs. We find the DSA-2000, CHORD, CASM, and CHIME/FRB will have effective optical depths that are 
2.8, 4, 16, and 16 times lower than BATSE GRBs, respectively.
This is because FRBs in those surveys will be more nearby than typical GRBs.
We find that, contingent on GRB\,950830 having been lensed, DSA-2000 will find 0.1--7.3 FRBs lensed by IMBHs after five years of observing at 80$\%$ efficiency.
Similarly, CHORD will find 0.1--7.0 events, assuming the BATSE GRB was lensed and the sample's mean redshift was roughly 2. CASM would detect 
0.3--7.2 and CHIME/FRB could expect 0.01--0.29 IMBH-lensed bursts. 

\subsection{Massive galaxy lenses}
FRBs lensed by massive galaxies will experience time-delays 
from days to months. The FRB's host galaxy should also 
be gravitationally lensed and will allow for 
modeling of the lensing galaxy's mass profile. 
Unlike time-variable non-transient sources like quasars, 
the FRB is only ``on'' for a millisecond and will 
not obscure the lensed host galaxy. Furthermore, 
FRBs lead to time-delay uncertainties that are more than 
ten orders of magnitude smaller than previous applications 
of time-delay cosmography. 
For these reasons, FRB lensing
by massive galaxies has been suggested as a new 
tool for measuring the Hubble parameter \citep{li2018}. \citet{Wucknitz2021} even proposed using gravitational lenses as 
a galaxy-scale interferometer to resolve structure and motion in the FRB itself. 
Most of these methods require repeating FRBs to break the mass-sheet degeneracy \citep{falco-1985}. Interferometric localization will alleviate the need for temporal coherence, as the 
lensed FRB images could be spatially resolved for typical 
image separations.

The probability of 
detecting a lensed event whose delay is longer than a survey 
pointing time or a transit time ($\sim$\,$10^3$\,s for DSA-2000, CHORD, and 
CHIME/FRB), is very low. If a telescope is always pointing at the same 
patch of sky, either because it is ultra-widefield (a CASM such as BURSTT) 
or through special survey strategy, it can in principle detect lensing 
delays up to the survey duration ($\sim$\,years). For this reason, 
we consider only CASM when forecasting detection rates of FRB lensing 
by massive galaxies. 

We use the lensing rate formalism presented in Equation~\ref{eq:lens-rate-eq} 
to estimate the number of FRBs lensed by massive galaxies. We take the 
magnification bias for an SIS dark matter halo and source power-law luminosity 
function with cumulative index $\gamma=1$. We find that 
after five years of observing, our putative CASM would 
detect 5--40 FRBs lensed by massive galaxies. 
The lower limit 
comes from assuming no magnification bias and 12,500 CASM
FRB detections per year. The upper bound assumes 
$\gamma=1.0$ and $B=4$ with 25,000 FRBs per year

\subsection{Probing the circumgalactic medium}

\begin{figure}[htb]
    \centering
    \includegraphics[width=0.8\columnwidth]{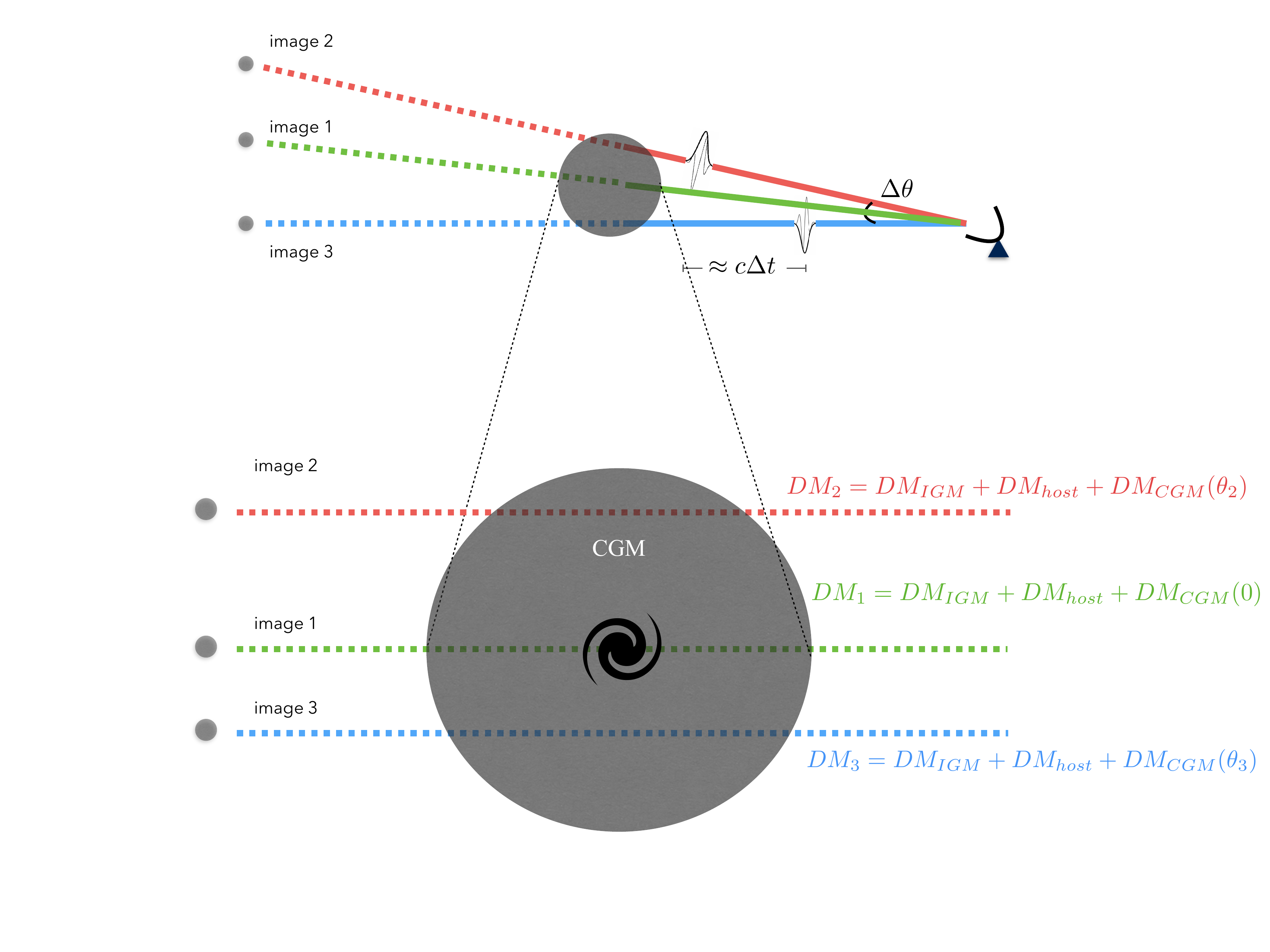}
    \caption{A schematic diagram of using gravitationally lensed FRBs 
    as probes of the circumgalactic medium of lensing galaxies. Each image 
    will have the same host DM, IGM DM, and contribution from the Milky Way, 
    so the only difference in observed DM between multiple copies will 
    be attributable to the lensing galaxy's CGM.}
    \label{fig:cgm-lens}
\end{figure}

Most proposed methods studying the  
circumgalactic medium (CGM) using FRBs have relied on 
the ensemble statistics of large numbers of sources, 
using DM and its line-of-sight 
statistics to study the distribution halo gas \citep{mcquinn2014, connorravi22}. 
Another approach is to model the contribution to DM 
from the host galaxy, the intergalactic medium (IGM), and 
the Milky Way to infer the component imparted by 
the halos of intervening galaxies \citep{Prochaska2019, ravi2019}. Both approaches are 
limited by modelling uncertainties. 
But an FRB lensed by a massive intervening galaxy will 
probe that galaxy's halo gas along multiple sight-lines. This would be a 
clean measure of that galaxy's CGM, as the only difference in propagation 
for the multiple paths will arise due to the galaxy lens. The method would 
constrain both the radial profile of halo gas density, as well as the 
CGM's inhomogeneity. A diagram of this idea is shown in 
Figure~\ref{fig:cgm-lens}.
\begin{figure}[htb]
    \centering
    \includegraphics[width=0.65\columnwidth]{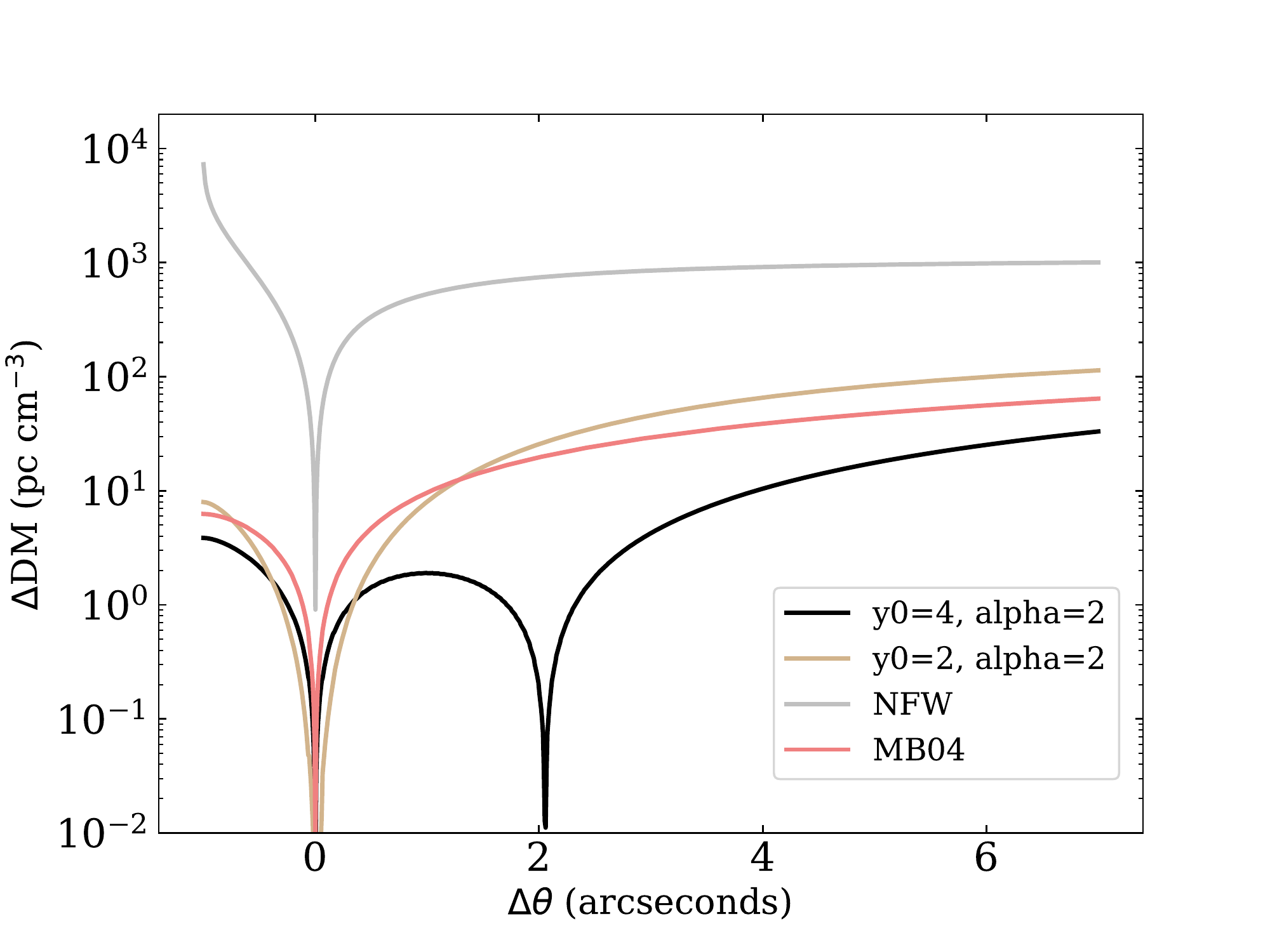}
    \caption{The observed difference in DM between two copies of 
    an FRB lensed by a $10^{12}$\,$M_\odot$ galaxy. $\Delta$DM 
    is plotted as a function of image separation, $\Delta\theta$, assuming the first image is at $\theta_1=1$''. We have used four simple, spherically symmetric models for the free electron distribution in the lensing 
    galaxy CGM. The $y_0=4$ curve has 
    two zero points because its DM curve 
    is non-monotonic.}
    \label{fig:cgm-dm}
\end{figure}

A similar technique has been used for many years in quasar absorption 
studies, where multiply-imaged quasars have been used to measure 
the structure and differential composition of the CGM of intervening 
lensing galaxies \citep{rauch-2002}, even on 
 spatial scales as small as 400\,pc \citep{Rudie-2019}.
However, such studies measure metal-bearing gas and 
cannot easily constrain the total baryonic matter in the halo, due to large uncertainties when extrapolating from 
the relatively rare metals to total gas content.

With strong lensing of an FRB detected with very long baseline interferometry (VLBI), 
the halo of 
the galaxy lens could be studied directly by comparing the 
DM, RM, and scattering properties of the multiple copies. Even if the 
lensing galaxy is a massive elliptical, \citet{zahedy-2019} has shown that such quiescent halos still have a rich CGM. We assume that the angular positions 
of the lensed copies can be measured to $\lesssim$0.25\,arcseconds with VLBI outriggers.
From this, one would have both the angular separation of the lenses images and 
the lensing time-delay. VLBI localization would give the angular impact parameter, $\theta$, allowing 
one to determine where in the intervening galaxy's halo the FRB passed through. 
The DM from an intervening halo will be given by the following integral,

\begin{equation}
    \mathrm{DM}_{CGM} = 2\,\int\displaylimits^{\sqrt{r^2_{max} - b^2}}_0 \frac{n_e(r)}{1+z_l}\,dr,
\end{equation}

\noindent where $r_{max}$ is a halo cutoff radius, 
which we will take to be the virial radius, $r_{200}$, 
and,

\begin{equation}
    b \approx \theta D_l.
\end{equation}

\noindent The observed quantity is the difference in DM between the two 
images. This is given by the difference in sightlines 
from the lensing galaxy's CGM, $\Delta \mathrm{DM} = \mathrm{DM}_{CGM}(\theta_1) - \mathrm{DM}_{CGM}(\theta_2)$ given by,

\begin{equation}
    \Delta \mathrm{DM} = \frac{2}{1+z_l} \int\displaylimits^{r_1}_{r_2} n_e(r)\,dr.
\end{equation}

\noindent Here we have written $\Delta\,\mathrm{DM}$ as a single definite integral with $r_1=\sqrt{r^2_{max} - \theta^2_1\,D^2_l}$ and $r_2=\sqrt{r^2_{max} - \theta^2_2\,D^2_l}$. At present, we do not know 
$n_e(r)$. But we will have measured $\Delta \mathrm{DM}$, the lens redshift, $z_l$, and 
$\theta_1$ and $\theta_2$. In Figure~\ref{fig:cgm-dm} we plot 
$\Delta \mathrm{DM}$ for a range of $\Delta \theta$, holding 
$\theta_1$ fixed at 1''. We use four models from \citet{prochaska2019b}, but this a non-exhaustive list. The first is a classic Navarro-Frenk-White (NFW) profile \citep{nfw1997}, which can be generalized to an 
modified NFW (mNFW) profile as,

\begin{equation}
    \rho(r) = \frac{\rho_b}{y^{1-\alpha}\,(y_0 + y)^{2+\alpha}},
\end{equation}

\noindent with $y \equiv c\frac{r}{r_{200}}$ where 
$r$ is radius, $r_{200}$ is the virial radius, and $c$ is a concentration parameter. For the standard NFW, $\alpha=0$ and
$y_0=1$. We also consider two mNFWs with ($y_0=2$, $\alpha=2$) and ($y_0=4$, $\alpha=2$). The final model, MB04, comes from \citet{maller2004}. The DMs and halo models 
were calculated with publicly available 
code\footnote{https://github.com/FRBs/FRB}.

The NFW profile is unrealistically steep at small radii,
leading to large DM differences even for small angular 
separations (see Figure~\ref{fig:cgm-dm}). The other three models are distinguishable 
for $\Delta\theta \geq 0.5''$, since observed 
FRB DM uncertainties with voltage data are often 
less than 0.5\,pc\,cm$^{-3}$. Combining $\Delta\,\mathrm{DM}$ with the difference in 
rotation measure, $\Delta\,$RM, will  
give a clean measure of the CGM line-of-sight 
magnetic field difference between the two 
impact parameters. This measurement will be 
valuable because the magnetic field in galaxy halos 
remains largely unconstrainted observationally \citep{vandefoort-2020}.

\section{Conclusions}

Fast radio bursts offer a unique probe of the 
Universe's dark matter via gravitational lensing, 
thanks to their abundance, short duration, and 
the coherent nature of their detection. In principle,
FRBs can give us access to $\sim$\,fifteen orders 
of magnitude in lens mass, corresponding to lensing 
time delays of microseconds to years. 
This is made possible by saving phase-preserving voltage data 
(the electric field waveform of the radio pulse itself). We have 
provided an overview of FRB gravitational lensing spanning a wide range of time-delays and lens masses.
We have forecasted detection rates for upcoming and current surveys. 
These include DSA-2000, CHORD, a coherent all-sky monitor (CASM) such as BURSTT, as well as CHIME/FRB, which is currently finding 
large numbers of FRBs and searching for lensing events. 
The FRB redshift distribution is critical 
to any lensing optical depth calculation because lensing 
probability is a strong function of source distance. 
This creates a trade-off between deep, sensitive telescopes 
such as DSA-2000 and CHORD, and ultra-widefield but less 
sensitive CASM surveys. 

On 1--100 microseconds timescales, FRBs will be 
lensed by stars in foreground galaxies at a significant 
rate and could plausibly be detected by CHIME/FRB.
If lensing by stars can be distinguished 
from compact dark objects, FRB surveys 
will limit the abundance of 
primordial black holes in the mass range 0.1--100\,$M_\odot$. 
This mass range is under-explored and is favoured by 
some theorists to be the most relevant mass scale for PBHs \citep{pbh-review2020}. 
All surveys we have considered will also constrain the 
cosmological density of intermediate mass black holes, 
$\Omega_{IMBH}$. If recent claims of GRB millilensing 
by IMBHs \citep{Paynter-2022} are correct, DSA-2000, CHORD, and the putative CASM survey could all detect a few lensed FRBs on timescales of 0.01--10\,s over the course of their surveys. On longer delay timescales 
(days to years), FRBs will be lensed by massive galaxies. 
Several cosmological applications have been proposed 
for these systems \citep{li2018,Wucknitz2021}.
Detecting strong lensing by galaxies is difficult because 
it is not known when the lensed copy will arrive, so the 
radio telescope must be pointing at the same location most of the time. Of the telescopes and survey strategies 
considered here, only a CASM-like survey (e.g. a BURSTT-like telescope but with a CHIME/FRB SEFD) could plausibly find 
FRB lensing by massive galaxies, with 5--40 such 
events after five years of observing. 

Finally, we have proposed a new method for studying the 
circumgalactic medium of lensing galaxies. Multiply imaged FRBs will traverse the lensing halos' CGM
at different impact parameters and will be 
differentially dispersed, Faraday rotated, or even scattered. Unlike 
with previous methods, this provides a clean model-independent measure of the total baryonic material 
in dark matter halos as a function of radius. Our method 
requires sub-arcsecond localization of each lensed 
copy of the FRB, which will only be possible 
for CASM-like surveys that have VLBI outriggers. 

\vspace{2cm}
\textbf{\noindent Data availability statement} This work was based on publicly available data 
and can be reproduced with the 
Jupyter notebook found in https://github.com/liamconnor/frb-grav-lensing.

\begin{acknowledgements}
We thank Sterl Phinney and Minghao Yue for valuable insights on strong lensing. We also thank Matt Dobbs, Keith Vanderlinde, Zarif Kader, and Ue-Li Pen for helpful discussions on coherent gravitational lensing.
\end{acknowledgements}

\bibliography{frb_grav_lensing}{}
\bibliographystyle{aasjournal}

\end{document}